\documentclass[
twocolumn,
aps,
prl,
reprint,
superscriptaddress,
amsmath,amssymb,
longbibliography
]{revtex4-2}
\usepackage[pdftex]{graphicx}
\usepackage[pdftex,
            colorlinks=true,
            citecolor=blue,
            linkcolor=blue]{hyperref}
\graphicspath{{./fig},{./fig_SM}}

\usepackage{braket}
\usepackage{times,multirow,amsfonts,bm,xspace,pifont}
\usepackage{soul}
\usepackage[normalem]{ulem}
\usepackage{textcomp}

\begin{document}
\newcommand{\sgn}{\mathrm{sgn}\,}
\newcommand{\tr}{\mathrm{tr}\,}
\newcommand{\Tr}{\mathrm{Tr}\,}
\newcommand{\mG}{\mathcal{G}}
\newcommand{\Log}{\mathrm{Log}\,}

\newcommand{\YY}[1]{\textcolor{magenta}{#1}}
\newcommand{\AD}[1]{\textcolor{blue}{#1}}
\newcommand*{\ADS}[1]{\textcolor{blue}{\sout{#1}}}
\newcommand*\YYS[1]{\textcolor{magenta}{\sout{#1}}}
\newcommand{\reply}[1]{{#1}}
\newcommand{\replyS}[1]{\textcolor{red}{\sout{#1}}}

\renewcommand{\Re}{\mathrm{Re}}
\renewcommand{\Im}{\mathrm{Im}}

\newcommand{\halpha}{{\hat{\alpha}}}
\newcommand{\uint}{{\int_0^\infty}}
\newcommand{\mA}{{\mathcal{A}}}
\newcommand{\mP}{{\mathcal{P}}}
\newcommand{\mB}{{\mathcal{B}}}
\newcommand{\mJ}{{\mathcal{J}}}
\newcommand{\bx}{{\bar{x}}}
\newcommand{\by}{{\bar{y}}}
\newcommand{\bz}{{\bar{z}}}
\newcommand{\bepsilon}{{\bar{\epsilon}}}

\newcommand{\SC}{{\rm{s}}}
\newcommand{\normal}{{\rm{n}}}

\newcommand{\Tc}{{T_{\rm c}}}
\newcommand{\qc}{{q_{\rm c}}}
\newcommand{\Tcn}{{T_{\rm c0}}}
\newcommand{\RK}{{R_{\rm K}}}

\newcommand{\header}[1]{{\textit{#1. }---}}

\title{
Quantum Theory of Current-Generating Local Orbital Magnetization
}

\author{Akito Daido} 
\email[]{daido@scphys.kyoto-u.ac.jp}
\affiliation{Department of Physics, Graduate School of Science, Kyoto University, Kyoto 606-8502, Japan}
\date{\today}

\begin{abstract}
Local orbital magnetization is the field whose rotation generates the equilibrium current density. Unlike spin magnetization, a quantum-mechanical local formula consistent with both this current relation and the modern theory of bulk orbital magnetization has been missing.
In this work, we derive a quantum-mechanical formula for the local orbital magnetization for non-interacting electrons by considering local-flux response of the grand potential.
The local-flux response fixes the formula uniquely in two dimensions, whereas in three dimensions it selects a natural representative within a longitudinal ambiguity.
Furthermore, coarse graining yields a natural local marker that generates the current to third-derivative order, and its site-position moment equals the orbital magnetic quadrupole moment of finite-size systems.
We illustrate the obtained results with the Haldane model.
\end{abstract}

\maketitle

\header{Introduction}
Magnetism is a fascinating phenomenon with continuing discoveries of novel magnetic states.
While many of the magnetic states are characterized by real-space spin alignment,
some are characterized by the orbital motion of electrons~\cite{Varma1997-ho,Chakravarty2001-iu,Resta2010-if}, 
as illustrated by recently discussed orbital ferromagnetic states in graphene-based heterostructures~\cite{Tschirhart2021-zh,Grover2022-mz,Han2023-en} and time-reversal-breaking charge order in kagome superconductors~\cite{Jiang2021-lj,Mielke2022-gs}. Considering that spin magnetism is classified and quantitatively characterized by real-space spin configurations,
it is natural to seek analogous description for orbital magnetism.
However, such description remains less developed, due to the lack of the established quantum-mechanical formulation of local orbital magnetization.

Local orbital magnetization is the orbital contribution to the field $\bm{m}(\bm{x})$ introduced through the magnetization current density $\bm{j}(\bm{x})=\nabla_{\bm{x}}\times\bm{m}(\bm{x})$.
As widely recognized, $\bm{m}(\bm{x})$ can be expressed by using $\bm{j}(\bm{x})$ 
up to a rotation-free field, e.g., $\bm{m}(\bm{x})=\nabla_{\bm{x}}\varphi+\int_1^\infty d\lambda\,\lambda\bm{x}\times{\bm{j}(\lambda\bm{x})}$~\cite{Hirst1997-ei}. It may be possible to define $\bm{m}(\bm{x})$ with some ``gauge fixing" such as $\varphi=0$, and define the local orbital magnetization by subtracting spin expectation value.
However, such expression is physically unsatisfactory.
Indeed, physical local magnetization should be insensitive to the system properties away from the reference point $\bm{x}$, while the $\varphi=0$ expression is not. Furthermore, the unit-cell average of the local orbital magnetization should reproduce the modern theory of bulk orbital magnetization~\cite{Xiao2010-ah,Xiao2005-px,Thonhauser2005-gi,Ceresoli2006-fs,Shi2007-rc,Lopez2012-ut,Zhu2012-hc,Nourafkan2014-ju}, but this is not obvious in the above. 
While there have been several efforts to define local orbital magnetization~\cite{Bianco2013-vb,Bianco2016-yk,Seleznev2023-tz,Saati2025-hg,Mahon2023-dk, Swiecicki2014-nb, Mahon2019-hd, Mahon2022-ye}, quantum-mechanical formulas satisfying all the physical requirements have not been known, lacking either clear connection to the defining relation $\bm{j}(\bm{x})=\nabla_{\bm{x}}\times\bm{m}(\bm{x})$~\cite{Bianco2013-vb,Bianco2016-yk,Seleznev2023-tz,Saati2025-hg}
or consistency with the bulk orbital magnetization in metals and Chern insulators~\cite{Mahon2023-dk, Swiecicki2014-nb, Mahon2019-hd, Mahon2022-ye}.

In this work, we show that it is indeed possible to construct a quantum-mechanical formula for local orbital magnetization satisfying the physical requirements.
Our strategy is summarized as follows.
We consider a simply-connected finite-size system with open boundary conditions (OBCs).
The current density can be expressed as ${j_i(\bm{x})}=-\delta\Omega/\delta A_i(\bm{x})$, i.e., the functional derivative of the grand potential $\Omega$ with respect to the vector potential $A_i(\bm{x})$.
On the other hand, the gauge invariance implies that the grand potential in normal metals and insulators responds to the vector-potential variation $\delta\bm{A}$ only through the magnetic field $\delta\bm{B}=\nabla\times\delta\bm{A}$ in this system geometry~\cite{Supplemental}.
Therefore, we introduce a local magnetization field $\bm{m}$ by $ \delta\Omega=-\int d^d\bar{x}\ \bm{m}(\bar{\bm{x}})\cdot\delta\bm{B}(\bar{\bm{x}})$ in $d=2,3$ dimensions.
Denoting the unit vector along the $x_i$ axis as $\hat{x}_i$, we obtain ${j_i(\bm{x})}
    =\int d^d\bar{x}\ \bm{m}(\bar{\bm{x}})\cdot\nabla_{\bar{\bm{x}}}\times[\hat{x}_i\delta(\bm{x}-\bar{\bm{x}})]=\epsilon_{ijk}\partial_{x_j}m_k(\bm{x})$. Thus, in contrast to previous approaches~\cite{Bianco2013-vb,Bianco2016-yk,Seleznev2023-tz,Saati2025-hg}, our construction automatically guarantees the exact relation to the equilibrium current density.

In the following, we microscopically perform this procedure for noninteracting electrons and derive local orbital magnetization formulas for both the continuum and lattice models. 
While we consider OBCs,
locality allows us to express the local orbital magnetization at bulk reference points by the quantities in periodic boundary conditions (PBCs), and the modern theory of bulk orbital magnetization~\cite{Xiao2010-ah,Shi2007-rc} is recovered by taking the unit-cell average.
OBCs also make the relation to the boundary magnetization and magnetic quadrupole moments (MQMs) transparent. 
The obtained results are illustrated with the Haldane model~\cite{Haldane1988-og}.

\header{Local magnetization in continuum models}
We start from the  local magnetization in continuum models,
with the Hamiltonian of the form
$H=\frac{(\hat{\bm{p}}-\bm{A}(\hat{\bm{r}}))^2}{2m}+V_1(\hat{\bm{r}})
    -\bm{B}(\hat{\bm{r}})\cdot\bm{\sigma}+(\hat{\bm{p}}-\bm{A}(\hat{\bm{r}}))\cdot\bm{\sigma}\times\nabla V_2(\hat{\bm{r}})-\mu$,
which can describe non-interacting electrons with spin-orbit interaction in solids coupled to vector potential.
With higher powers of $\hat{\bm{p}}-\bm{A}(\hat{\bm{r}})$ or additional explicit magnetic-field coupling, model-dependent local magnetization terms may appear
but can be included in the same local-flux-response framework~\cite{Supplemental}.
The grand potential $\Omega$ can be expressed as
    $\Omega=-T\sum_{\omega_n}e^{i\omega_n(+0)}\int_{-\infty}^0 du\sum_\alpha\int d^dr\,\braket{\bm{r}\alpha|G_u|\bm{r}\alpha}$,
since 
it vanishes for the chemical potential $\mu\to-\infty$ and thus is given by integrating $\partial\Omega/\partial\mu$ from $-\infty$ to $\mu$.
Here, we defined the Green's function $G_u\equiv(i\omega_n+u-H)^{-1}$, Matsubara frequency $\omega_n=2\pi T(n+1/2)$ with temperature $T$, 
and the position eigenstate by $\hat{\bm{r}}\ket{\bm{r}\alpha}=\bm{r}\ket{\bm{r}\alpha}$ with the spin state $\alpha=\uparrow,\downarrow$.

Note that the grand potential $\Omega$ is expressed by the local matrix elements of the Green's function.
Importantly, it is known that perturbation to such local matrix elements can be 
directly expressed by that of the magnetic flux~\cite{Chen2011-dj}.
After evaluating the variation [See End Matter~\cite{EndMatter}], we obtain $\delta\Omega=-\int d^d\bar{x}\,[\bm{m}_{\rm orb}(\bar{\bm{x}})+\bm{m}_{\rm spin}(\bar{\bm{x}})]\cdot\delta\bm{B}(\bar{\bm{x}})$ and thus
\begin{gather}
{\bm{j}(\bm{x})}=\nabla_{\bm{x}}\times[\bm{m}_{\rm orb}(\bm{x})+\bm{m}_{\rm spin}(\bm{x})],\label{eq:j_m_relation}
\end{gather}
by considering a local variation of the vector potential. 
Here, $m^i_{\rm spin}(\bm{x})$ is the expectation value of the spin density operator, and
the orbital contribution takes the three-point form
\begin{subequations}\label{eq:3pts}\begin{align}
    m^i_{\rm orb}(\bm{x})&=\int_{\bm{r}_1\bm{r}_2\bm{r}_3}\,\chi_{\bm{r}_1\bm{r}_2\bm{r}_3}^i(\bm{x})\,m_{\bm{r}_1\bm{r}_2\bm{r}_3}\label{eq:formula_m_chi},
\end{align}
with $\int_{\bm{r}_1\bm{r}_2\bm{r}_3}\equiv\int d^dr_1\int d^dr_2\int d^dr_3$. We defined a geometric factor
\begin{align}
    &\chi^i_{\bm{r}_1\bm{r}_2\bm{r}_3}(\bm{x})\equiv\int_{\bar{\bm{x}}\in\triangle_{123}}dS_i\,\delta(\bm{x}-\bar{\bm{x}}),\label{eq:chi_def}
\end{align}    
whose surface integral is taken over the triangle $\triangle_{123}$ formed by $\bm{r}_1$, $\bm{r}_2$, and $\bm{r}_3$;
For example, in two dimensions, 
this gives a unit vector either parallel or antiparallel to the out-of-plane direction for $\bm{x}$ inside the triangle $\triangle_{123}$ and vanishes elsewhere.
We also defined the three-point magnetization kernel
\begin{align}
    \! m_{\bm{r}_1\bm{r}_2\bm{r}_3}\equiv-i\sum_{mn}\frac{F_{mn}}{\epsilon_{mn}}\Tr[n(\bm{r}_1)Hn(\bm{r}_2)p_m n(\bm{r}_3)p_n],
\end{align}\end{subequations}
to which the system properties are encoded.
We defined the eigenvalue $\epsilon_{m}$ of the Hamiltonian $H$ and the projector to its eigenspace $p_m$, and 
the energy-dependent quantities 
$\epsilon_{mn}\equiv\epsilon_m-\epsilon_n$, $F_{mn}\equiv F(\epsilon_m)-F(\epsilon_n)$, and $F(\epsilon)\equiv-T\ln(1+e^{-\epsilon/T})$~\footnote{For $\epsilon_m=\epsilon_n$, $F_{mn}/\epsilon_{mn}\equiv\partial_{\epsilon_m}F(\epsilon_m)$}.
We also defined the projection to the position $\bm{r}$ by $n(\bm{r})\equiv\sum_\alpha\ket{\bm{r}\alpha}\bra{\bm{r}\alpha}$, which also plays the role of the charge-density operator.
Green's function formula is available in End Matter [Eq.~\eqref{eq:formula_m_all}].

The magnetization field given in Eq.~\eqref{eq:3pts}
has the desired properties for physical local orbital magnetization.
Indeed, Eq.~\eqref{eq:3pts} exactly satisfies the defining relation~\eqref{eq:j_m_relation}.
Moreover, Eq.~\eqref{eq:3pts} depends on system properties only around the reference point $\bm{x}$ 
at finite temperature  or in insulators with an energy gap, since
the integrals over the positions $\bm{r}_1$, $\bm{r}_2$, and $\bm{r}_3$ are contributed only around $\bm{x}$ owing to the spatial decay of Green's functions [See End Matter].
This also indicates that Eq.~\eqref{eq:3pts} is insensitive to the boundary conditions when $\bm{x}$ is away from the boundaries.
For such bulk reference points,
we can show that unit-cell average of Eq.~\eqref{eq:3pts} reproduces the quantum-mechanical formula for the bulk orbital magnetization~\cite{Thonhauser2005-gi,Shi2007-rc} [See Supplemental Material~\cite{Supplemental}].
We will call Eq.~\eqref{eq:3pts} the three-point formula in the following.

Note that in three dimensions, the current-generating property, or the procedure obtaining $\bm{m}_{\rm orb}(\bm{x})$ from $\delta\Omega$, leaves a longitudinal ambiguity 
as discussed in End Matter~\cite{EndMatter}.
Thus, the three-point formula should be understood as a physically natural representative selected by the local-flux-response construction.
On the other hand, such ambiguity is reduced to a constant in two dimensions and can be removed by the physical criteria in the above.
Thus, in two dimensions, we can call the three-point formula the local orbital magnetization.

\header{Local orbital magnetization in lattice models}
Next, we consider a lattice system described by a Hamiltonian of the form
$H=\sum_{\bm{r}\bm{r}'\alpha\beta}
\ket{\bm{r}\alpha}[H_{\bm{r}\bm{r}'}]_{\alpha\beta}
\bra{\bm{r}'\beta}$.
Here, $\ket{\bm{r}\alpha}$ denotes a basis state localized at the
real-space position $\bm{r}$, where $\bm{r}$ includes both the
unit-cell and sublattice positions, while $\alpha$ labels the
remaining internal degrees of freedom.
We introduce the vector potential through the Peierls substitution
$H_{\bm{r}\bm{r}'}=[H_{\bm{A}=0}]_{\bm{r}\bm{r}'}
e^{iA_{\bm{r}\bm{r}'}}$,
where
$A_{\bm{r}\bm{r}'}\equiv
\int_{\bm{r}'}^{\bm{r}}d\bar{\bm{x}}\cdot\bm{A}(\bar{\bm{x}})$
is evaluated along the straight line from $\bm{r}'$ to $\bm{r}$.
This vector-potential coupling defines the microscopic
current distribution considered below.
Explicit magnetic-field couplings such as the Zeeman term can be straightforwardly
included as in the
continuum case.
In the following, we focus on the orbital contribution to the
magnetization, while $H$ may have e.g., nontrivial spin dependence.

In lattice models, the current operator is defined on each bond rather than lattice site, and the local current density $j_i(\bm{x})$ is not defined.
To make the discussion parallel to the continuum model, we interpolate the lattice quantities to continuous space by defining the smeared charge-density operator $\bar{n}(\bm{x})\equiv\sum_{\bm{r}}W(\bm{x}-\bm{r})n(\bm{r})$ at any spatial position $\bm{x}$, which can be off the lattice sites.
Here, the smearing function $W(\bm{x})$ is arbitrary as long as $\int d^dx\,W(\bm{x})=1$ is satisfied: For example, by choosing $W(\bm{x})$ to be the delta function, $\bar{n}(\bm{x})$ reflects the tightly-bound charge configuration, while representing the coarse-grained charge density by choosing $W$ to be a macroscopic envelope function.
We can show that the smeared current-density operator $\bar{j}_i(\bm{x})$ consistent with $\bar{n}(\bm{x})$ can be defined by considering the response of $H$ to the vector-potential variation $\delta\bm{A}(\bm{r})\propto\hat{x}_iW(\bm{x}-\bm{r})$.
Indeed, $\bar{j}_i(\bm{x})$ defined this way satisfies the continuity equation $i[H,\bar{n}(\bm{x})]+\nabla_{\bm{x}}\cdot\bar{\bm{j}}(\bm{x})=0$ and reproduces the current operator $J_i\equiv i[H,\hat{r}_i]$ after integrated over space~\cite{EndMatter}.

Now the derivation of the local magnetization formula is parallel to that of continuum models.
We obtain the smeared current density in the form of magnetization current
$\braket{\bar{{j}}_i(\bm{x})}=\epsilon_{ijk}\partial_{x_j}\bar{{m}}^k_{\rm orb}(\bm{x})$, with the smeared magnetization 
\begin{align}
    \bar{{m}}^i_{\rm orb}(\bm{x})=\int d^d\bar{x}\,W(\bm{x}-\bar{\bm{x}})\,{m}^i_{\rm orb}(\bar{\bm{x}})\label{eq:mbar_def}.
\end{align}
Here, ${m}^i_{\rm orb}(\bm{x})$ is given by the three-point formula Eqs.~\eqref{eq:3pts} and \eqref{eq:formula_m_all} with replacing the position integrals $\int_{\bm{r}_1\bm{r}_2\bm{r}_3}$ with the site summation $\sum_{\bm{r}_1\bm{r}_2\bm{r}_3}$.
Thus, we can define the local orbital magnetization by the three-point formula in lattice models as well, in that it reproduces the smeared current density and magnetization for arbitrary smearing function $W$.

\begin{figure}
    \centering
    \includegraphics[width=1\linewidth]{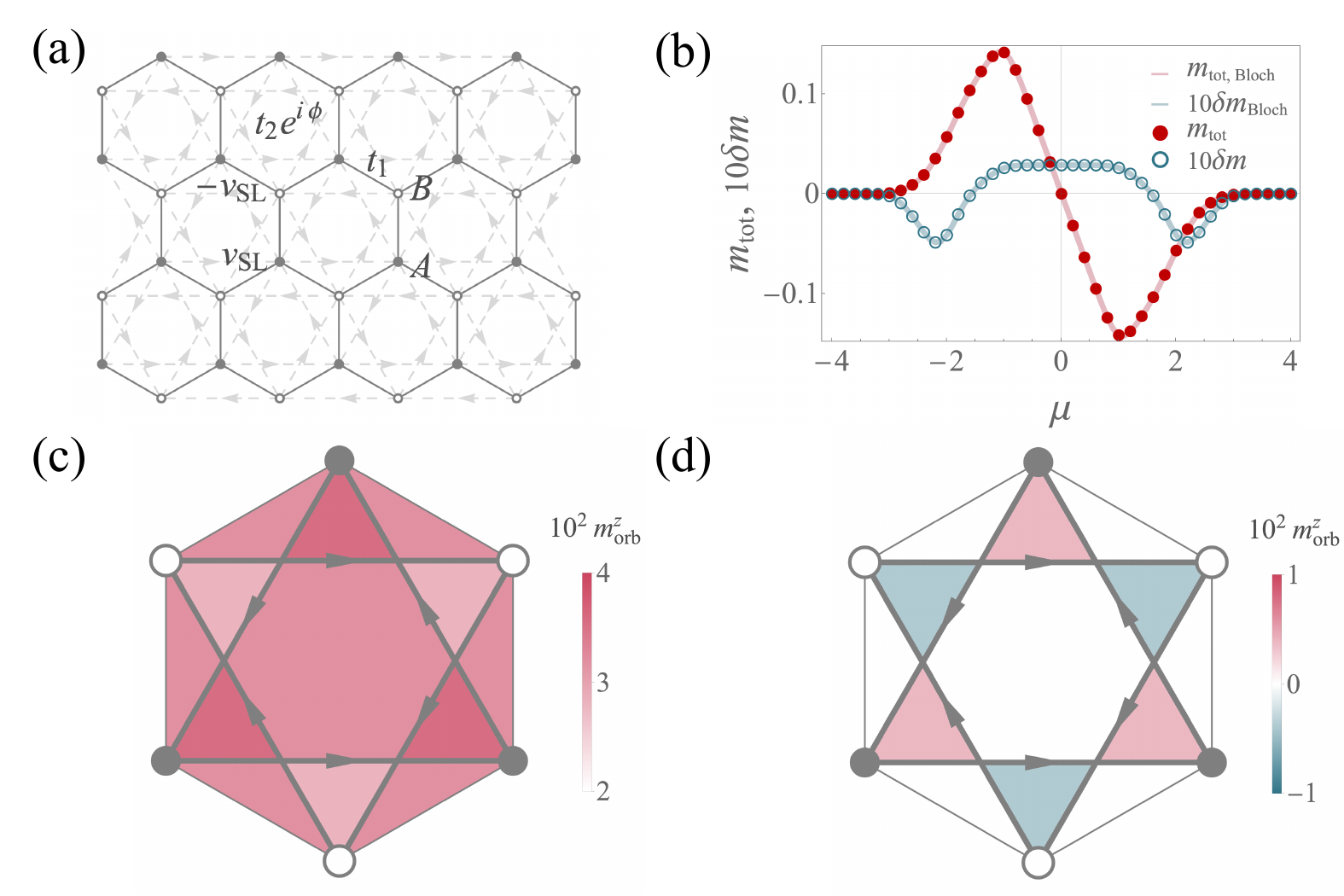}
    \caption{(a) Haldane model with zigzag and armchair edges
    illustrated for $N_{\rm site}=36$. 
    (b) Chemical potential $\mu$ dependence of NLM at bulk lattice sites evaluated with OBCs and the Bloch-state formulas. 
    Here, $m_{\rm tot}=(\mathcal{M}_z(A)+\mathcal{M}_z(B))/V_{\rm UC}$ (red) and $10\delta m\equiv10(\mathcal{M}_z(A)-\mathcal{M}_z(B))/V_{\rm UC}$ (blue), with the unit-cell area $V_{\rm UC}=3\sqrt{3}a^2/2$ and the nearest-neighbor bond length $a=1$. 
    (c), (d) Three-point formula in a bulk unit cell for (c) $\mu=-0.2$ and (d) $\mu=0$ shown with bond currents (gray arrows). 
    }
    \label{fig1}
\end{figure}

We can establish a clear relationship between the three-point formula $m^i_{\rm orb}(\bm{x})$ and the bond-current distribution of lattice models. By choosing $W$ to be the delta function,  $m^i_{\rm orb}(\bm{x})$ itself determines the smeared current density by $\braket{\bar{j}_i(\bm{x})}=\epsilon_{ijk}\partial_{x_j}m^k_{\rm orb}(\bm{x})$.
This can be associated with the bond current by integrating over an infinitesimally small area $S_{\rm fi}$ threaded by the bond $\bm{r}_{\rm i}\to\bm{r}_{\rm f}$, and we obtain the identity
\begin{align}
    \oint_{\partial S_{\rm fi}} d\bm{x}\cdot\bm{m}_{\rm orb}(\bm{x})=\braket{J_{\bm{r}_{\rm i}\to\bm{r}_{\rm f}}},\label{eq:m_and_Jbond}
\end{align}
connecting the three-point formula with the expectation value of the bond-current operator $J_{\bm{r}_{\rm i}\to\bm{r}_{\rm f}}\equiv-\partial_{A_{\bm{r}_{\rm f}\bm{r}_{\rm i}}}H$~\cite{Supplemental}.
In two dimensions, this means that the local orbital magnetization discretely changes when the reference point $\bm{x}$ crosses a bond $\bm{r}_{\rm i}\to\bm{r}_{\rm f}$ by its bond current $\braket{J_{\bm{r}_{\rm i}\to\bm{r}_{\rm f}}}$ and otherwise remains constant.
In other words, two-dimensional lattice models can be viewed as a collection of microscopic ferromagnetic domains with the domain walls set by the bond currents [See Fig.~\ref{fig1}(c),(d) for example].

\header{Coarse graining and the NLM}
While the three-point formula gives a magnetization field exactly generating the current density, it may not be practically useful as it involves three site summations and detailed information smaller than the lattice constant.
This motivates us to simplify the formula with keeping only the contributions that are essential after coarse graining.
To be specific, we focus on lattice models below, while the expressions for continuum models are obtained by replacing site summations with position integrals.

We consider a smearing function $W$ that slowly varies in space with a length scale $l_W$.
Note that by performing the $\bar{\bm{x}}$ integral in Eq.~\eqref{eq:mbar_def}, the position dependence of $\bar{m}^i_{\rm orb}(\bm{x})$ is determined by
integrating $W(\bm{x}-\bar{\bm{x}})$ for $\bar{\bm{x}}\in\triangle_{123}$ through the geometric factor Eq.~\eqref{eq:chi_def}.
When we write $\bar{\bm{x}}=\bm{r}_1+\delta\bar{\bm{x}}$,
the displacement $\delta\bar{\bm{x}}$ is as small as some microscopic length scale $l$, since the triangle $\triangle_{123}$ has to be small for the magnetization kernel to be non-negligible owing to its locality.
This allows the Taylor expansion 
     $W(\bm{x}-\bar{\bm{x}})=\sum_{n=0}^\infty
     \frac{1}{n!}(-\delta\bar{\bm{x}}\cdot\nabla_{\bm{x}})^nW(\bm{x}-\bm{r}_1)$, with the small parameter $\delta\bar{\bm{x}}\cdot\nabla_{\bm{x}}\sim l/l_W$.
Repeating the same procedure for the other two vertices of the triangle $\triangle_{123}$ and taking the average, we can show that 
\begin{align}
\braket{\bar{j}_i(\bm{x})}=\epsilon_{ijk}\partial_{x_j}[\bar{\mathcal{M}}_k(\bm{x})+\partial_{x_a}\partial_{x_b}\delta\bar{\mathcal{M}}_{kab}(\bm{x})],\label{eq:local_marker_current}
\end{align}
for some field $\delta\bar{\mathcal{M}}_{kab}(\bm{x})$.
Here we defined $\bar{\mathcal{M}}_i(\bm{x})\equiv\sum_{\bm{r}}W(\bm{x}-\bm{r})\mathcal{M}_i(\bm{r})$
and 
\begin{align}
    \mathcal{M}_i(\bm{r})&=\frac{\epsilon_{ijk}}{6}\sum_{mn}\frac{F_{mn}}{\epsilon_{mn}}\,\Tr\Bigl[n(\bm{r})\Bigl([\hat{r}_j,p_m]\,p_nJ_k\notag\\
    &\quad+J_k[\hat{r}_j,p_m]\,p_n+p_nJ_k[\hat{r}_j,p_m]\Bigr)\Bigr]
    \label{eq:m1}.
\end{align}
Equation~\eqref{eq:local_marker_current} indicates that the quantity ${\mathcal{M}}_i(\bm{r})$, after coarse graining, correctly reproduces the coarse-grained current density
up to a cubic total derivative.

As understood from the derivation, $\mathcal{M}_i(\bm{r})$ is the redistribution of the local orbital magnetization to lattice sites,
with keeping the coarse-grained magnetization approximately unchanged. Thus, $\mathcal{M}_i(\bm{r})$ naturally defines the orbital magnetic moment associated to each site $\bm{r}$, rather than every spatial point $\bm{x}$ as the three-point formula~\eqref{eq:3pts} does.
Note also that $\mathcal{M}_i(\bm{r})$ has the form of the so-called local marker~\cite{Bianco2011-tv,Bianco2013-vb,Bianco2016-yk,Seleznev2023-tz}, a magnetization-related site-wise quantity.
There have been several proposals of the local markers, by requiring the sum rule $M_{\rm orb}^i=\sum_{\bm{r}}\mathcal{M}_i(\bm{r})$ with the total magnetic moment $M_{\rm orb}^i\equiv- \partial\Omega/\partial B_i$ as the guiding principle~\cite{Bianco2013-vb,Bianco2016-yk,Seleznev2023-tz}, but there still remains ambiguity in its definition~\cite{Seleznev2023-tz}.
In this regard, $\mathcal{M}_i(\bm{r})$ has a clear advantage and 
provides a natural choice of the local marker: It not only satisfies the sum rule but also gives a good approximation of the coarse-grained current density $\braket{\bar{j}_i(\bm{x})}$, 
whereas markers constrained only by the sum rule need not control $\braket{\bar{j}_i(\bm{x})}$ beyond lower-order gradient terms.
We will call Eq.~\eqref{eq:m1} the natural local marker (NLM) for this reason. 
The Bloch-state formula for the NLM in a bulk lattice point is available in Supplemental Material.

\header{Orbital magnetization and quadrupole moment}
We next consider the application of the three-point formula and NLM to the orbital MQM. This is defined with OBCs by 
$M_{\rm orb}^{ij}\equiv\frac{1}{3}\int d^dx\,x_i\left(\bm{x}\times\braket{\bar{\bm{j}}(\bm{x})}\right)_j,$
whose symmetric part, traceless by construction, describes the far-field configuration of the magnetic field.
This can also be expressed as a $W$-independent expression
$M_{\rm orb}^{ij}=\frac{1}{12}\braket{\{\hat{r}_i,\epsilon_{jkl}\{\hat{r}_k,J_l\}\}}.$ Here and hereafter, we assume that the smearing function $W(\bm{x})$ is isotropic.

By using the relation $\braket{\bar{j}_i(\bm{x})}=\epsilon_{ijk}\partial_{x_j}\bar{m}_{\rm orb}^k(\bm{x})$,
the orbital MQM can be expressed based on the smeared magnetization by
$M_{\rm orb}^{ij}=\int d^dx\,x_i\bar{m}_{\rm orb}^j(\bm{x})-\delta_{ij}x_k\bar{{m}}_{\rm orb}^k(\bm{x})/3$.
This immediately leads to 
a MQM expression based on the three-point formula,
    $M_{\rm orb}^{ij}=\int d^dx\,x_i{m}_{\rm orb}^j(\bm{x})-\frac{\delta_{ij}}{3}x_k{{m}}_{\rm orb}^k(\bm{x}),$
by choosing $W$ to be the delta function.
Similar formulas for higher-order multipole moments can also be obtained in the same way. 
Furthermore, in the case of the MQM, 
we can use the NLM instead of the three-point formula $m^i_{\rm orb}(\bm{x})$, since the  $\delta\bar{\mathcal{M}}$ term in Eq.~\eqref{eq:local_marker_current} does not contribute after integration by parts.
We thus obtain
\begin{align}
    M_{\rm orb}^{ij}
    &=\sum_{\bm{r}}r_i\mathcal{M}_j(\bm{r})-\frac{\delta_{ij}}{3}r_k\mathcal{M}_k(\bm{r}).\label{eq:Qij_formula}
\end{align}
The fact that orbital MQM is obtained as the site-position moment of $\mathcal{M}_j(\bm{r})$ supports the interpretation of NLM as the site-wise orbital magnetic moment.

The MQM formulas in the above are equivalent to each other regardless of the system size, and can be applied not only to macroscopic materials but also to small molecular systems.
Their equivalence can also be directly proven~\cite{Supplemental}.
The MQM formulas remain valid even with a net orbital magnetic moment $M^i_{\rm orb}$, while $M^{ij}_{\rm orb}$ has an origin dependence in this case.
In summary, the three-point formula and NLM establish the quantum-mechanical counterpart of the equivalence between current- and magnetization-based magnetic multipole moments in classical electromagnetism.

\header{Illustration in the case of the Haldane model}
We illustrate the obtained formulas in the case of the Haldane model~\cite{Haldane1988-og} on the OBC flake illustrated in Fig.~\ref{fig1}(a).
Unless otherwise noted, we focus on the topological phase with $t_1=1$, $t_2=0.5t_1$, $\phi=\pi/2$, $v_{\rm SL}=0.3t_1$, and $T=0.1t_1$, by using the number of sites $N_{\rm site}=3566$. Model details are available in Supplemental Material~\cite{Supplemental}.

We first show the chemical potential $\mu$ dependence of the NLM in Fig.~\ref{fig1}(b), where red and blue circles correspond to the summation and difference of NLM on the two sublattices $A$ and $B$ located away from the boundaries.
They are reproduced with the Bloch-state formulas as shown by the solid lines, demonstrating the boundary insensitivity of the NLM in the bulk.
We can see that the net magnetic moment vanishes for $\mu=0$ and therefore the quadrupole moment should be dominant.
In the following, we focus on $\mu=-0.2$ and $\mu=0$ and call them the orbital ferromagnet (OFM) and the orbital quadrupole magnet (OQM) states, respectively.

Figures~\ref{fig1}(c) and (d) show the three-point formula $m^z_{\rm orb}(\bm{x})$ inside a bulk unit cell for the OFM and OQM states.
We can see that the average of the local orbital magnetization is consistent with the bulk orbital magnetization $m_{\rm tot}$ in Fig.~\ref{fig1}(b).
We can also see that the bond currents run consistently with those expected from $\epsilon_{ijk}\partial_{x_j}m_{\rm orb}^k(\bm{x})$.
In particular, the discontinuity of $m_{\rm orb}^z(\bm{x})$ satisfies Eq.~\eqref{eq:m_and_Jbond}; for example, $|m_{\rm orb}^z|$ is negligibly small in the white regions of Fig.~\ref{fig1}(d) and $|m_{\rm orb}^z|= 3.8\times 10^{-3}$ in all the red and blue regions as ensured by the (bulk) threefold rotation and rotation-particle-hole symmetries~\cite{Supplemental}, while all the next-nearest-neighbor bond currents have the magnitude $|\braket{J_{AA}}|=|\braket{J_{BB}}|= 3.8\times 10^{-3}$ and the nearest-neighbor bond currents are negligibly small.

\begin{figure}
    \centering
    \includegraphics[width=\linewidth]{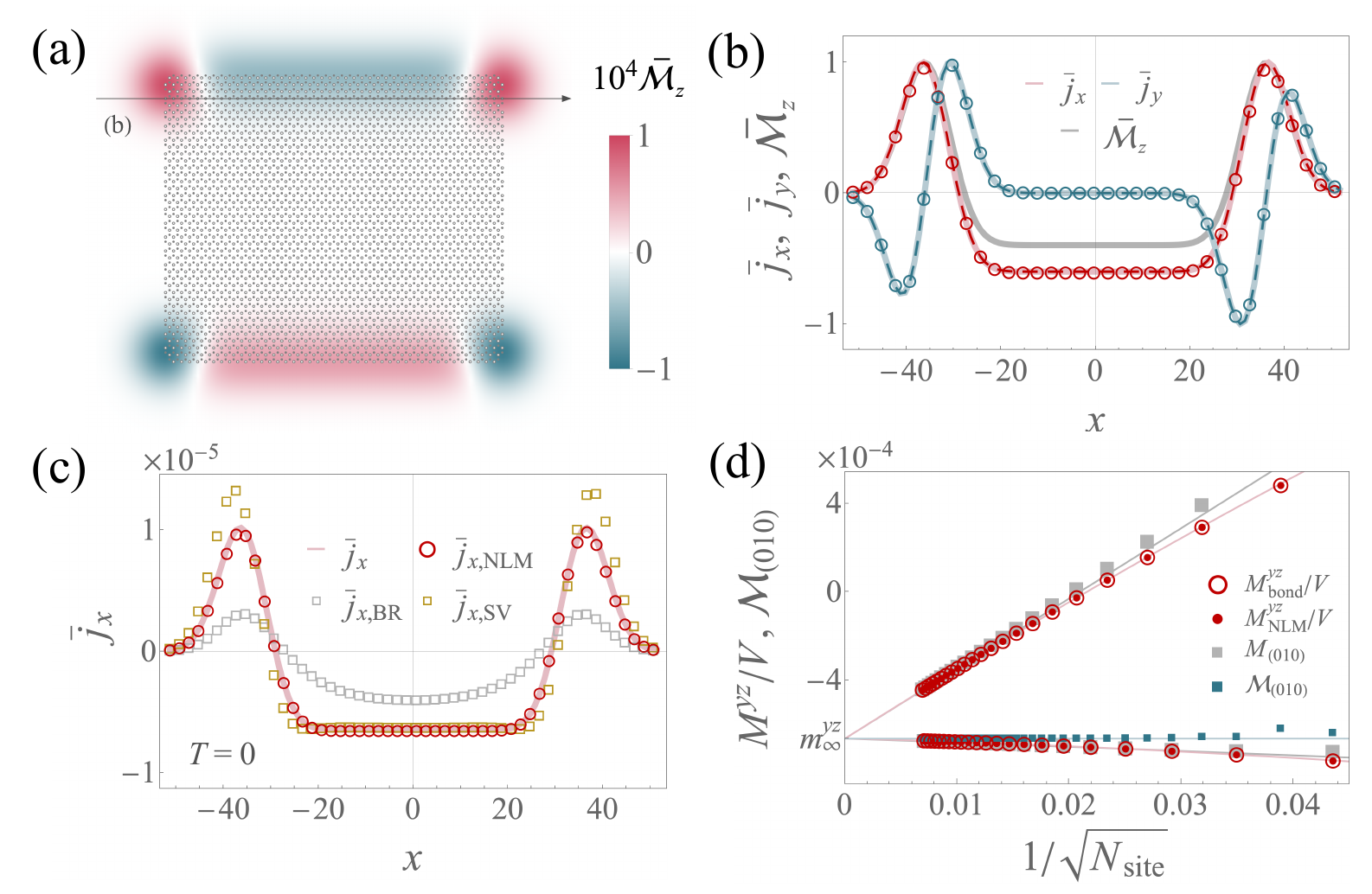}
    \caption{
    Edge magnetization, current density, and MQM in the OQM state.
(a) Gaussian-smeared NLM $\bar{\mathcal{M}}_z$.
(b) Smeared currents on the line in (a) from bond currents, three-point formula, and NLM, with $\bar{\mathcal{M}}_z$ in gray, normalized to $[-1,1]$~\cite{Supplemental}.
(c) $T=0$ comparison of $\bar j_x$ with BR and SV markers.
(d) Size scaling of $M^{\rm orb}_{yz}/V$ from bond currents and NLM, with averaged edge moment and edge magnetization. $V$ represents the system area. The two branches reflect different corner terminations~\cite{Supplemental}.
    }
    \label{fig2}
\end{figure}

Below we focus on the OQM state and show the profile of the magnetization and current density smeared by the Gaussian $W(\bm{x})\equiv\frac{1}{2\pi l_W^2}e^{-|\bm{x}|^2/2l_W^2}$ with $l_W=5a$. The smeared NLM $\bar{\mathcal{M}}_z(\bm{x})$ of the OQM state in Fig.~\ref{fig2}(a) shows edge magnetization on the $(010)$ and $(0\bar{1}0)$ edges with sign reversals near the corners, while the total magnetic moment vanishes as ensured by rotation-particle-hole symmetry~\cite{Supplemental}.
As shown in Fig.~\ref{fig2}(b) for along the line indicated in Fig.~\ref{fig2}(a),
the magnetization current expected from the NLM [open circles], i.e., the first term of Eq.~\eqref{eq:local_marker_current}, qualitatively reproduces the exact smeared current density evaluated from the bond currents [light red/blue solid lines] even though $l_W/a=\mathcal{O}(1)$.
The three-point formula [dashed lines] agrees with the bond-current result to numerical precision,
as expected from their analytical equivalence.

To make comparison with local markers in previous studies, we show the result for $T=0$ in Fig.~\ref{fig2}(c).
The magnetization current densities defined from the Bianco-Resta (BR)~\cite{Bianco2013-vb} and Seleznev-Vanderbilt linear-combination (SV)~\cite{Seleznev2023-tz} local markers do not capture the exact smeared current density, while they agree with the NLM result for the trivial phase of the Haldane model [See Fig.~\ref{fig3}(a)].
Similarly, the finite-$T$ marker proposed in Ref.~\cite{Saati2025-hg} does not capture the current density of the OQM state~\cite{Supplemental}.
These 
results show that markers constrained by sum rules alone do not necessarily satisfy the additional local current-generating criterion.

As expected from Fig.~\ref{fig2}(a), the OQM state has finite MQM $M^{yz}_{\rm orb}$, as shown in Fig.~\ref{fig2}(d) for various values of $N_{\rm site}$.
MQM evaluated by NLM [Eq.~\eqref{eq:Qij_formula}] coincides with the MQM evaluated by the bond currents
for arbitrary system sizes, supporting their analytical equivalence.
We also showed the averaged edge magnetic moment $M_{(010)}\equiv\frac{1}{D_x}\int_{-\infty}^\infty dx\int_{0}^\infty dy\,\bar{\mathcal{M}}_z(\bm{r})$ with a system diameter in the $x$ direction $D_x$~\cite{Supplemental}, and edge magnetization $\mathcal{M}_{(010)}$ near the edge center $x=0$ [See Eq.~\eqref{eq:mathcalM_surf} in End Matter].
The edge magnetization immediately converges to a value $m^{yz}_\infty=-6.6\times 10^{-4}$, as expected from the corner insensitivity, and the other quantities also do in the thermodynamic limit.
Comparison with the previously-proposed local markers is available in End Matter and Supplemental Material~\cite{Supplemental}.

\header{Discussion}
The local-flux formulation promotes orbital magnetization to a current-generating real-space field 
on the same footing as spin magnetization. It provides a microscopic basis for orbital MQMs and surface orbital magnetization, including 
hinge-current phenomena in higher-order topological phases~\cite{Zhu2021-bo,Seleznev2023-tz,Gliozzi2022-gg}.
The formulas may also be useful for real-space descriptions of orbital magnetism and orbitronics~\cite{Jo2024-le,Cysne2025-ra}.

\begin{acknowledgments}
This work was supported by JSPS KAKENHI (Grant No.JP23K17353, No.JP23KK0248, and No. JP26H00659).
The use and verification of AI-assisted code generation are described in the Supplemental Material.
\end{acknowledgments}

%

\newpage

\appendix
\begin{center}\textbf{End Matter}\end{center}
In this work, we set $\hbar=1$ and $-|e|\bm{A}(\bm{x})$ is replaced with $\bm{A}(\bm{x})$ for simplicity.
The electron charge $|e|$ and the Dirac constant $\hbar$ can be recovered by multiplying $-|e|/\hbar$ to all the magnetization-related formulas.

\section{Derivation of three-point formula}
We show how to evaluate the variation of $\Omega$.
We denote the Green's function for $\bm{A}(\bm{r})+\delta\bm{A}(\bm{r})$ as $\tilde{G}_u$. Then, we can show the identity for its inverse,
\begin{align}
    \braket{\bm{r}\alpha|\tilde{G}^{-1}_u|\bm{r}'\beta}=\braket{\bm{r}\alpha|{G}_u^{-1}+\delta\bm{B}(\hat{\bm{r}})\cdot\bm{\sigma}|\bm{r}'\beta}e^{i\delta A_{\bm{r}\bm{r}'}}.\label{eq:Identity3}
\end{align}
Here, 
$\delta A_{\bm{r}\bm{r}'}\equiv\int_{\bm{r}'}^{\bm{r}}d\bar{\bm{x}}\cdot\delta\bm{A}(\bar{\bm{x}})$ is the Peierls-phase variation.
Equation~\eqref{eq:Identity3} clearly holds in lattice models and is derived for the continuum model in Supplementary Materials~\cite{Supplemental,Essin2010-jr,Luttinger1951-hj}.

We write $[G_u]_{\bm{r}\bm{r}'}$ as the matrix in spin space. 
Then, following Ref.~\cite{Chen2011-dj}, the Dyson equation for $\tilde{G}_u$ reads, with $e^{i\delta A_{\bm{r}\bm{r}}}=1$ and introducing $[\mG_u]_{\bm{r}_2\bm{r}_3}=e^{-i\delta A_{\bm{r}_2\bm{r}_3}}[\tilde{G}_u]_{\bm{r}_2\bm{r}_3}$,
\begin{align}
    \delta(\bm{r}_1-\bm{r}_3)&=\int d^dr_2[\tilde{G}_{u}^{-1}]_{\bm{r}_1\bm{r}_2}[\tilde{G}_{u}]_{\bm{r}_2\bm{r}_3}e^{i\delta A_{\bm{r}_3\bm{r}_1}}\label{eq:temp4}\\
    &=\int d^dr_2[G_u^{-1}+\delta\bm{B}(\hat{\bm{r}})\cdot\bm{\sigma}]_{\bm{r}_1\bm{r}_2}[\mG_u]_{\bm{r}_2\bm{r}_3}e^{i\delta\phi_{123}}.\notag
\end{align}
Here, we defined the flux threading the triangle $\triangle_{123}$ 
formed by $\bm{r}_1$, $\bm{r}_2,$ and $\bm{r}_3$ by $\delta\phi_{123}\equiv\delta A_{\bm{r}_1\bm{r}_2}+\delta A_{\bm{r}_2\bm{r}_3}+\delta A_{\bm{r}_3\bm{r}_1}$, i.e.,
\begin{align}
\delta\phi_{123}
&=\int_{\bar{\bm{x}}\in\triangle_{123}}d\bm{S}\cdot\delta\bm{B}(\bar{\bm{x}}),
\end{align}
with defining the orientation of $\triangle_{123}$ accordingly.
This shows that, in the OBC geometry considered here, $\mathcal{G}_u$ depends on the vector-potential perturbation only via gauge-invariant triangle fluxes, or equivalently via the magnetic-field variation $\delta\bm B$.

Now, the three-point formula can straightforwardly be obtained. By using $[\tilde{G}_u]_{\bm{r}\bm{r}}=[\mG_u]_{\bm{r}\bm{r}}$, the change of the grand potential $\delta\Omega$
can be described by the trace of $\mG_u-G_u$, which can be obtained by expanding the Dyson equation in terms of $\delta\bm{B}$.
The current density is obtained by $-\lim_{\varepsilon\to0}\delta\Omega/\varepsilon$ with substituting $\varepsilon\hat{x}_i\delta(\bm{x}-\bar{\bm{x}})$ for $\delta\bm{A}(\bar{\bm{x}})$ in the above.
This leads to 
the spin magnetization ${m}^i_{\rm spin}(\bm{x})=T\sum_{\omega_n}e^{i\omega_n\eta}\Tr[{\sigma}^in(\bm{x})G]$ with $\eta=+0$ and\begin{subequations}\label{eq:formula_m_all}\begin{align}
    {m}^i_{\rm orb}(\bm{x})&=iT\sum_{\omega_n}e^{i\omega_n\eta}\int^0_{-\infty}du\int_{\bm{r}_1\bm{r}_2\bm{r}_3}I^i_{123},\label{eq:formula_m}
\end{align}
with
\begin{align}
\!\! I^i_{123}\equiv    \Tr\Bigl[n(\bm{r}_1)G_u^{-1}n(\bm{r}_2)G_un(\bm{r}_3)G_u\Bigr]{\chi}^i_{\bm{r}_1\bm{r}_2\bm{r}_3}(\bm{x}).\label{eq:morb_GF}
\end{align}\end{subequations}
The three-point formula~\eqref{eq:3pts} is obtained after taking the Matsubara summation and the $u$ integral. 
A step-by-step derivation is illustrated 
in Supplemental Material.

\section{Uniqueness and longitudinal ambiguity of the orbital magnetization}
Abbreviating the spin contribution for simplicity, the three-point formula~\eqref{eq:3pts} is defined to satisfy
\begin{equation}
    \delta\Omega=-\int d^d\bar{x}\ \bm{m}_{\rm orb}(\bar{\bm{x}})\cdot \delta\bm{B}(\bar{\bm{x}}),\label{eq:dOmegadB}
\end{equation}
for any infinitesimal magnetic-field variation $\delta\bm{B}$ generated by a vector-potential variation, and hence satisfying $\nabla\cdot\delta\bm{B}=0$. 
In three dimensions, this constraint prevents us from choosing 
$\delta\bm B(\bar{\bm{x}})\propto \hat{x}_i\delta(\bm{x}-\bar{\bm{x}})$ as required to define a functional derivative $-\delta\Omega/\delta B_i(\bm{x})$.
Consequently, Eq.~\eqref{eq:dOmegadB} does not identify all components of $\bm{m}_{\rm orb}$ and leaves a longitudinal ambiguity.
Indeed, for any smooth scalar function $\varphi'$ that vanishes outside the sample, we obtain 
$\int d^3\bar{x}\,\nabla\varphi'\cdot\delta\bm{B}=0.$
Thus $\bm{m}_{\rm orb}$ and $\bm{m}_{\rm orb}+\nabla\varphi'$ give the same $\delta\Omega$ and the same current density.
Therefore, in three dimensions, the three-point formula should be regarded as a representative magnetization field 
selected by the triangle-flux construction rather than as a unique magnetization field, while it provides a natural representative consistent with the physical requirements.

The situation is different in two dimensions, and the three-point formula $m^z_{\rm orb}(\bm{x})=-\delta\Omega/\delta B_z(\bm{x})$ uniquely gives the local orbital magnetization consistent with both the current condition $\bm{j}(\bm{x})=\nabla_{\bm{x}}\times\bm{m}_{\rm orb}(\bm{x})$ and the modern theory of bulk orbital magnetization.
Indeed, the current density in two dimensions is generated by a scalar function $m^z_{\rm orb}(\bm{x})$ with $\nabla_{\bm{x}} m^z_{\rm orb}(\bm{x})=\hat{z}\times\bm{j}(\bm{x})$,
which indicates that difference of two magnetization fields generating the same current density is a constant.
This constant can be fixed by requiring either $m^z_{\rm orb}(\bm{x})$ to vanish outside of the system; $\int d^2x\,m^z_{\rm orb}(\bm{x})$ to reproduce the total magnetic moment; or the unit-cell average of $m^z_{\rm orb}(\bm{x})$ in periodic crystals to recover the modern theory of orbital magnetization.
The three-point formula satisfies all of them, and is unique in this sense.

A local extension of the thermodynamic relation~\cite{Xiao2010-ah} follows in two and three dimensions in the same sense. Taking the $\mu$ derivative of Eq.~\eqref{eq:dOmegadB} with $N=-\partial\Omega/\partial\mu$, we obtain 
$\delta N=\int d^dx\, \partial_\mu\bm{m}_{\rm orb}(\bm{x})\cdot\delta\bm{B}(\bm{x})$ for all allowed magnetic-field variations.
In three dimensions, this relation fixes the part of $\partial_\mu\bm{m}_{\rm orb}(\bm{x})$ that couples to the divergence-free field $\delta\bm{B}(\bm{x})$, while
in two dimensions, it gives the identity $\partial m^z_{\rm orb}(\bm{x})/\partial\mu=\delta N/\delta B_z(\bm{x})$.

\section{Locality of the three-point formula}
Here we show that the three-point formula at $\bm{x}$ is contributed by only around the reference point $\bm{x}$.
For $T>0$, Eq.~\eqref{eq:3pts} can be rewritten as
\begin{align}
    m^i_{\rm orb}(\bm{x})
    &=-i\sum_{\bm{r}_1\bm{r}_2\bm{r}_3}\chi^i_{\bm{r}_1\bm{r}_2\bm{r}_3}(\bm{x})\oint_{C}\frac{dz}{2\pi i}\Tr\Bigl[n(\bm{r}_1)Hn(\bm{r}_2)\notag\\
    &\qquad\qquad\cdot \left.\frac{1}{z-H}n(\bm{r}_3)\frac{F(z)-F(H)}{z-H}\right].\label{eq:formula_m_T}
\end{align}
The integration contour is set to $C=(-\infty-i\frac{\pi T}{2},\infty-i\frac{\pi T}{2})+(\infty+i\frac{\pi T}{2},-\infty+i\frac{\pi T}{2})$,
by using the analyticity of the integrand.
Considering a short-ranged Hamiltonian $H$, $1/(z-H)$ decays in space as is the case for Matsubara Green's function. Therefore, $\bm{r}_1$, $\bm{r}_2$, and $\bm{r}_3$ have to be close to each other, and thus to $\bm{x}$ as well, to make a non-negligible contribution to $m^i_{\rm orb}(\bm{x})$.
This can also be inferred from Eq.~\eqref{eq:morb_GF} by noting that the Matsubara Green's function decays in space.

We can also express the three-point formula for insulators at $T=0$ as $ m^i_{\rm orb}(\bm{x})=\sum_{\bm{r}_1\bm{r}_2\bm{r}_3}\chi^i_{\bm{r}_1\bm{r}_2\bm{r}_3}(\bm{x})m_{\bm{r}_1,\bm{r}_2,\bm{r}_3},$
with
\begin{align}\label{eq:formula_m_ins}
m_{\bm{r}_1,\bm{r}_2,\bm{r}_3}&=-\Im \oint_{\Re z<0}\frac{dz}{2\pi i}\Tr\left[\{H,n(\bm{r}_1)Hn(\bm{r}_2)\}\right.\notag\\
&\qquad\cdot\left.\frac{P}{z-H}n(\bm{r}_3)\frac{Q}{z-H}\right],
\end{align}
where $P$ and $Q$ are the projectors to the negative and positive energy states, respectively.
The contour integral is taken to surround the negative part of the real axis and ensure
the spatial decay of $1/(z-H)$ due to the finite band gap.
Thus, we also conclude the locality of the three-point formula.
Note that the present discussion relies on the presence of the spectral gap and thus an algebraic decay is expected near the boundaries with topological boundary states connecting bulk valence and conduction bands. 

At $T=0$ in clean metals, the above exponential-locality argument should be replaced by the algebraic decay. Even in this case, the three-point formula gives a current-generating magnetization field in systems with OBCs.

\section{Smeared current density and continuity equation}
Here we discuss the details of the smeared current density in lattice models.
The magnetic field couples to the Hamiltonian by $H=\sum_{\bm{r}\bm{r}'}n(\bm{r})H_{\bm{A}=0}e^{iA_{\bm{r}\bm{r}'}}n(\bm{r}')$, since $\sum_{\bm{r}}n(\bm{r})=1$.
By changing the vector potential by $\delta\bm{A}(\bm{r})=\varepsilon\hat{x}_iW(\bm{x}-\bm{r})$, 
we define the smeared current-density operator by $\bar{j}_i(\bm{x})\equiv-\lim_{\varepsilon\to0}(H|_{\bm{A}\to\bm{A}+\delta\bm{A}}-H)/{\varepsilon}$, i.e.,
\begin{align}
    \bar{j}_i(\bm{x})&=-\sum_{\bm{r}\bm{r}'}n(\bm{r})Hn(\bm{r}')i\int_{\bm{r}'}^{\bm{r}}d\bar{\bm{x}}\cdot\hat{x}_iW(\bm{x}-\bar{\bm{x}}).
    \label{eq:barj_TB}
\end{align}
This coincides with the $W$-smearing of $-\delta H/\delta\bm{A}(\bm{\bar{x}})$.
Furthermore, after integrating Eq.~\eqref{eq:barj_TB} over space, we reproduce the current operator
${J}_i=\int d^dx\,\bar{j}_i(\bm{x})
    =i[H,\hat{\bm{r}}],$ by using $\int d^dx\,W(\bm{x})=1$.
The continuity equation can easily be verified by taking the divergence of the smeared current density operator:
\begin{align}
    \partial_{x_i}\bar{j}_i(\bm{x})&=-\sum_{\bm{r}\bm{r}'}n(\bm{r})Hn(\bm{r}')i\int^{\bm{r}}_{\bm{r}'}d\bar{\bm{x}}\cdot\nabla_{\bm{x}}W(\bm{x}-\bar{\bm{x}})\notag\\
    &=\sum_{\bm{r}\bm{r}'}n(\bm{r})Hn(\bm{r}')i[W(\bm{x}-\bm{r})-W(\bm{x}-\bm{r}')]\notag\\
    &=i[\bar{n}(\bm{x}),H].
\end{align}
Thus, the smeared current density and the smeared charge density operators satisfy the continuity equation.
For the use in the main text, we also define the bond current operator for the bond $\bm{r}_i\to\bm{r}_f$ by $J_{\bm{r}_i\to\bm{r}_f}\equiv -\partial_{A_{\bm{r}_f\bm{r}_i}}H$, i.e.,     $J_{\bm{r}_i\to\bm{r}_f}
    =-in(\bm{r}_f)Hn(\bm{r}_i)+in(\bm{r}_i)Hn(\bm{r}_f)$,
with $A_{\bm{r}_i\bm{r}_f}=-A_{\bm{r}_f\bm{r}_i}$.

\begin{figure}
    \centering
\includegraphics[width=1\linewidth]{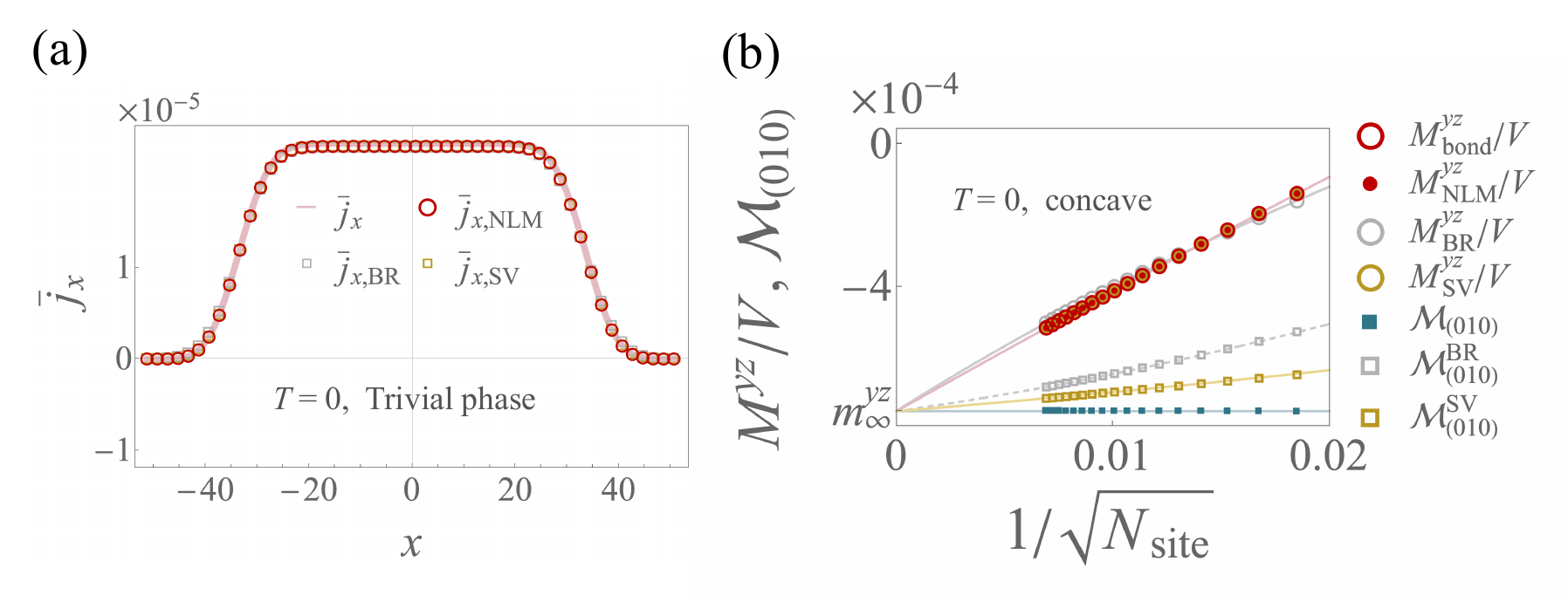}
    \caption{(a) The smeared current density at $T=0$ in the trivial phase along the same line as Fig.~\ref{fig2}(b,c). (b) The MQM and edge magnetization of the OQM state at $T=0$. The results only for the concave-type corner condition~\cite{Supplemental} are shown for simplicity.
    }
    \label{fig3}
\end{figure}

\section{Further comparison with BR- and SV-type markers}

To make further comparison of NLM with the BR- and SV-type markers, we show in Fig.~\ref{fig3}(a) the current-density distribution in the trivial phase at $T=0$ {($t_2=0.1$ and $v_{\rm SL}=2.0$)}.
All the magnetization currents defined from the NLM and BR- and SV- markers qualitatively agree with each other.
Figures~\ref{fig2}(c) and~\ref{fig3}(a) highlight that the sum-rule-based markers do not necessarily generate the correct coarse-grained current density in the presence of topological edge states.

We also show in Fig.~\ref{fig3}(b) the MQM-related quantities of the OQM state at $T=0$, while $T>0$ comparison of NLM with Ref.~\cite{Saati2025-hg} is discussed in Supplemental Material.
[Note that we are focusing on finite-system MQMs in OBCs, rather than thermodynamic bulk MQM densities in periodic crystals~\cite{Gao2018-cx,Shitade2018-lo}.]
All of them seem to converge to a value $m_{\infty}^{yz}=-7.5\times 10^{-4}$ in the thermodynamic limit. Notably, 
the MQM defined from either the NLM [filled circle] or the SV marker [open yellow circle]  coincides with the exact MQM [open red circle] at any finite-size systems. 
This is consistent with the numerical observations that the SV marker can predict the hinge currents and MQMs of finite-size system~\cite{Seleznev2023-tz}. 
The edge orbital magnetization 
\begin{align}
    \mathcal{M}_{(010)}\equiv\frac{1}{\sqrt{3}a}\sum_{\bm{r};x\sim0}\int_{0}^\infty dy\,\frac{e^{-\frac{(y-r_y)^2}{2l_W^2}}}{\sqrt{2\pi l_W^2}}{\mathcal{M}}_z(\bm{r}),\label{eq:mathcalM_surf}
\end{align}
is also shown,
where $\sum_{\bm{r};x\sim0}$ is taken over $-\sqrt{3}a/2<r_x\le\sqrt{3}a/2$, i.e., a translation unit along the $(010)$ edge.
The surface orbital magnetization defined from the NLM~\eqref{eq:mathcalM_surf} [filled square] showed a faster convergence to $m^{yz}_{\infty}$ than BR- and SV-markers [open squares], which are defined by replacing $\mathcal{M}_z(\bm{r})$ in Eq.~\eqref{eq:mathcalM_surf} with those by BR- and SV-markers.

\newpage
\begin{center}\textbf{Supplemental Material}\end{center}

\appendix

\section{Inverse of Green's function in continuum models}
Here we derive the identity Eq.~\eqref{eq:Identity3} for continuum models.
Note that we can write an operator identity\begin{subequations}\label{eq:U_db}\label{eq:19}\begin{align}
    \hat{\bm{p}}-\delta\bm{A}(\hat{\bm{r}})&=U_{\delta\bm{A}}(\hat{\bm{p}}-\delta\hat{\bm{b}}_{\bm{r}_c})U_{\delta\bm{A}}^\dagger,
\end{align}
with $U_{\delta\bm{A}}\equiv e^{i\delta A_{\hat{\bm{r}}\bm{r}_c}}$ and
\begin{align}
    \delta\hat{\bm{b}}_{\bm{r}_c}&=-\int_0^1dt\,t\left(\hat{\bm{r}}-\bm{r}_c\right)\times\delta\bm{B}\left({\bm{r}}_c+t\left(\hat{{\bm{r}}}-{\bm{r}_c}\right)\right),
\end{align}\end{subequations}
by following Ref.~\cite{Luttinger1951-hj}.
According to this identity, $\tilde{H}\equiv H_{\bm{A}\to\bm{A}+\delta\bm{A}}$ satisfies
\begin{align}
    \tilde{H}&=U_{\delta\bm{A}}H_{\bm{A}\to \bm{A}+\delta\bm{b}_{\bm{r}_c}}U^\dagger_{\delta\bm{A}}
    =U_{\delta\bm{A}}(H+\delta H)U^\dagger_{\delta\bm{A}},
\end{align}
with
\begin{align}
    \!\!\!\delta H&=-\frac{\hat{\bm{J}}\cdot\delta\hat{\bm{b}}_{\bm{r}_c}+\delta\hat{\bm{b}}_{\bm{r}_c}\cdot\hat{\bm{J}}}{2}+\frac{\delta\hat{\bm{b}}_{\bm{r}_c}^2}{2m}-\delta\bm{B}(\hat{\bm{r}})\cdot\bm{\sigma},\label{eq:dH_def}
\end{align}
and the current operator $  \hat{\bm{J}}=\frac{\hat{\bm{p}}-\bm{A}(\hat{\bm{r}})}{m}+\bm{\sigma}\times\nabla V_2(\hat{\bm{r}})$.

Let us consider the position matrix elements, particularly $\braket{\bm{r}\alpha|\delta H|\bm{r}'\beta}$.
We use Eq.~\eqref{eq:19} with 
setting $\bm{r}_c=(\bm{r}+\bm{r}')/2$.
Suppressing the spin indices $\alpha$ and $\beta$ for simplicity, we can show
\begin{align}
    \braket{\bm{r}|\hat{\bm{p}}\cdot\delta\hat{\bm{b}}_{\bm{r}_c}|\bm{r}'}=\braket{\bm{r}|\delta\hat{\bm{b}}_{\bm{r}_c}\cdot\hat{\bm{p}}|\bm{r}'}=0,
\end{align}
and $\braket{\bm{r}|\delta\hat{\bm{b}}_{\bm{r}_c}|\bm{r}'}=0$ by using, for example,
\begin{align}
    \epsilon_{ijk}\braket{\bm{r}|p_i(\hat{r}_j-r_{cj})|\bm{r}'}
    &=-\frac{1}{2}\epsilon_{ijk}\braket{\bm{r}|p_i|\bm{r}'}(r_j-r'_j)\notag\\
    &=-\frac{1}{2}\epsilon_{ijk}\braket{\bm{r}|[\hat{r}_j,\hat{p}_i]|\bm{r}'}\notag\\
    &=0,
\end{align}
and $\braket{\bm{r}|\delta\hat{\bm{b}}_{\bm{r}_c}|\bm{r}'}=\delta(\bm{r}-\bm{r}')\delta\hat{\bm{b}}_{\bm{r}_c}|_{\hat{\bm{r}}\to\bm{r}}=0$.
Accordingly, the matrix elements of the first two terms of Eq.~\eqref{eq:dH_def} vanish.
Furthermore, we obtain $\delta A_{\bm{r}\bm{r}_c}-\delta A_{\bm{r}'\bm{r}_c}=\delta A_{\bm{r}\bm{r}'}$ since $\bm{r}_c$ is chosen to be the midpoint of $\bm{r}$ and $\bm{r}'$.
Thus, we obtain
\begin{align}
    [\tilde{H}]_{\bm{r}\bm{r}'}&=[H-\delta \bm{B}(\hat{\bm{r}})\cdot\bm{\sigma}]_{\bm{r}\bm{r}'}e^{i\delta A_{\bm{r}\bm{r}'}}.\label{eq:24_SM}
\end{align}
This is a generalization of the similar relation proposed for the spatially uniform magnetic field~\cite{Essin2010-jr}.

The Green's-function identity used in End Matter,
\begin{align}
\braket{\bm{r}\alpha|\tilde{G}^{-1}_u|\bm{r}'\beta}=\braket{\bm{r}\alpha|{G}_u^{-1}+\delta\bm{B}(\hat{\bm{r}})\cdot\bm{\sigma}|\bm{r}'\beta}e^{i\delta A_{\bm{r}\bm{r}'}},\label{eq:8_SM}
\end{align}
immediately follows from Eq.~\eqref{eq:24_SM}
for the continuum Hamiltonian considered in this work.
For continuum Hamiltonians with higher powers of the momentum
or additional explicit magnetic-field couplings,
extra local $O(\delta B)$ terms generally appear in the square bracket of
Eq.~\eqref{eq:24_SM}.
Such model-dependent terms can be included in the same local-flux
response framework but are not part of the minimal continuum model considered here.

\section{Details of the derivation of the three-point formula}
In this section, we show a step-by-step derivation of the three-point formula for the local orbital magnetization in lattice models.
In the following, we consider a simply-connected finite-size system with OBCs.
In this case, as we can explicitly see below, the grand potential indeed depends on the vector potential only through the magnetic field.
This makes contrast to the PBCs or multiply-connected systems, where the grand potential can also depend on the Aharonov-Bohm fluxes $\Phi_{\rm AB}\equiv\oint d\bm{x}\cdot \bm{A}(\bm{x})$ along noncontractible loops. 
At fixed $T>0$, or in gapped systems, such winding contributions are
negligible for sufficiently large systems.
Thus, the boundary insensitivity is recovered in the thermodynamic limit as is also discussed below.

\subsection{Smeared current density}
We consider a general tight-binding Hamiltonian ${H}=\sum_{\bm{r}\bm{r}'\alpha\beta}\ket{\bm{r}\alpha}[H_{\bm{r}\bm{r}'}]_{\alpha\beta}\bra{\bm{r}'\beta}$.
Here, the state $\ket{\bm{r}\alpha}$ specifies a local orbital at spatial position $\bm{r}$, which includes the sublattice position, if any, with $\alpha$ the internal degrees of freedom other than sublattices such as spin.
The vector potential couples to the Hamiltonian through the Peierls substitution and the Zeeman coupling,
\begin{align}
    H_{\bm{r}\bm{r}'}=\Bigl([H_0]_{\bm{r}\bm{r}'}-\bm{B}(\bm{r})\cdot\bm{\sigma}\delta_{\bm{r}\bm{r}'}\Bigr)e^{iA_{\bm{r}\bm{r}'}},
\end{align}
where $H_0\equiv [H]_{\bm{A}=0}$ and $A_{\bm{r}\bm{r}'}$ is defined by
\begin{align}
    A_{\bm{r}\bm{r}'}\equiv\int^{\bm{r}}_{\bm{r}'}d\bar{\bm{{x}}}\cdot\bm{A}(\bar{\bm{x}}),
\end{align}
with the vector potential $\bm{A}(\bm{r})$.
The line integral is defined on the straight line connecting $\bm{r}'$ and $\bm{r}$.
The chemical potential $\mu$ is included to the Hamiltonian $H_0$.

To define the smeared current-density operator, we remember the case of continuum models. In this case, the microscopic current-density operator at position $\bm{x}$ is defined by $-\frac{\delta H}{\delta\bm{A}(\bm{x})}$, and thus
the smeared current-density operator is defined by 
\begin{align}
\bar{j}_i(\bm{x})=-\int d^d\bar{x}\,W(\bm{x}-\bar{\bm{x}})\frac{\delta H}{\delta\bm{A}(\bar{\bm{x}})}, \label{eq:barj_cm_SM}   
\end{align}
in $d$ dimensions
by using an arbitrary smearing function $W$
satisfying the normalization condition $\int d^d\bar{x}\,W(\bar{\bm{x}})=1$.
An alternative way to express $\bar{j}_i(\bm{x})$ is 
\begin{align}
    \bar{{j}}_i(\bm{x})&=-\lim_{\varepsilon\to0}\frac{H|_{\bm{A}\to\bm{A}+\delta \bm{A}}-H}{\varepsilon},\label{eq:coarse-grained current density}
\end{align}
where the variation of the vector potential is
\begin{align}
\delta\bm{A}(\bar{\bm{x}})\equiv\varepsilon\,\hat{x}_iW(\bm{x}-\bar{\bm{x}}),
\end{align}
with $\hat{x}_i$ the unit vector along the $x_i$ axis.

We use Eq.~\eqref{eq:coarse-grained current density} to define the smeared current density in tight-binding models at an arbitrary spatial position $\bm{x}$ not limited to lattice sites.
The obtained smeared current density operator can also be reproduced by formally evaluating $-\delta H/\delta\bm{A}(\bar{\bm{x}})$ in lattice models and substituting it for Eq.~\eqref{eq:barj_cm_SM}.
As shown in the End Matter, we can show that the smeared current density operator $\bar{\bm{j}}(\bm{x})$ satisfies the continuity equation 
\begin{align}
    [iH,\bar{n}(\bm{x})]+\nabla_{\bm{x}}\cdot\bar{\bm{j}}(\bm{x})=0,
\end{align}
with the smeared charge density
\begin{align}
    \bar{n}(\bm{x})&\equiv\sum_{\bm{r}}W(\bm{x}-\bm{r})n(\bm{r}),
\end{align}
which is defined on any point $\bm{x}$ in real space.
Here, the microscopic charge density operator on site $\bm{r}$ is defined by
\begin{align} n(\bm{r})\equiv\sum_\alpha\ket{\bm{r}\alpha}\bra{\bm{r}\alpha}.
\end{align}

The smeared current density at position $\bm{x}$ in equilibrium at temperature $T$ is given by
\begin{align}
    \braket{\bar{j_i}(\bm{x})}&=\frac{1}{\beta}\sum_{\omega_n}e^{i\omega_n\eta}\Tr[\bar{j}_i(\bm{x})G],
\end{align}
with the Green's function
$   G=(i\omega_n-H)^{-1}.$
Here $\omega_n=2\pi (n+1/2)/\beta$ with $n\in\mathbb{Z}$ is the Matsubara frequency, $1/\beta=T$ is the temperature, and $e^{i\omega_n\eta}$ with $\eta=+0$ is the convergence factor.
By using the continuity equation, we obtain the charge conservation
    $\nabla\cdot\braket{\bar{j}_i(\bm{x})}=0,$
which ensures the presence of the magnetization field $\bar{\bm{m}}(\bm{x})$ such that 
\begin{align}
    \braket{\bar{\bm{j}}(\bm{x})}=\nabla_{\bm{x}}\times\bar{\bm{m}}(\bm{x}).
\end{align}
In the following, we explicitly derive the quantum-mechanical formula to give the smeared magnetization $\bar{\bm{m}}(\bm{x})$.

It is convenient to consider the grand potential of the system defined by 
\begin{align}
    \Omega\equiv-\frac{1}{\beta}\Tr\Log(1+e^{-\beta H}).
\end{align}
By using $\Omega(\mu\to-\infty)=0$, we can rewrite $\Omega$ as
\begin{align}
    \Omega
    &=\Omega(\mu\to-\infty)+\int_{-\infty}^\mu d\mu'\,\frac{\partial\Omega}{\partial\mu'}\notag\\
    &=-\frac{1}{\beta}\sum_{\omega_n}e^{i\omega_n\eta}\int_{-\infty}^0 du\,\Tr G_u,
\end{align}
with
\begin{align}
G_u\equiv
(i\omega_n+u-H)^{-1}.
\end{align}
By using the grand potential, the smeared current density can be obtained by
\begin{align}
    \braket{\bar{j}_i(\bm{x})}
    &=-\lim_{\varepsilon\to0}\frac{1}{\varepsilon}\frac{1}{\beta}\sum_{\omega_n}e^{i\omega_n\eta}\Tr[\delta HG]\notag\\
    &=-\lim_{\varepsilon\to0}\frac{1}{\varepsilon}\delta\Omega,\label{eq:cj}
\end{align}
with $\delta H\equiv H_{\bm{A}\to \bm{A}+\delta\bm{A}}-H$ and
\begin{align}
    \delta\Omega\equiv \Omega|_{\bm{A}\to \bm{A}+\delta\bm{A}}-\Omega.\label{eq:deltaOmega}
\end{align}
Thus, we focus on the evaluation of 
\begin{subequations}
\begin{align}
    \delta\Omega&=-\frac{1}{\beta}\sum_{\omega_n}e^{i\omega_n\eta}\int_{-\infty }^0du\Tr[\tilde{G}_u-G_u],\label{eq:deltaOmega_40a}\\
    \tilde{G}_u&\equiv [G_u]_{\bm{A}\to\bm{A}+\delta\bm{A}},
\end{align}
\end{subequations}
to evaluate the equilibrium smeared current density.

\subsection{Variation of Green's function}
To evaluate the change of the grand potential upon $\delta\bm{A}$, we first derive that of Green's function.
It should be noted that
\begin{align}
    [\tilde{G}^{-1}_{u}]_{\bm{r}\bm{r}'}&=\left(i\omega_n+u+[\bm{B}(\bm{r})+\delta\bm{B}(\bm{r})]\cdot\bm{\sigma}\right)\delta_{\bm{r}\bm{r}'}\notag\\
    &\qquad-[H_0]_{\bm{r}\bm{r}'}e^{iA_{\bm{r}\bm{r}'}+i\delta A_{\bm{r}\bm{r}'}}\notag\\
    &=[G^{-1}_{u}-\delta H_{\rm Z}]_{\bm{r}\bm{r}'}e^{i\delta A_{\bm{r}\bm{r}'}},
\end{align}
by using $e^{i\delta A_{\bm{r}\bm{r}}}=1$ and the variation of the Zeeman coupling
\begin{align}
    \delta H_{\rm Z}\equiv-\sum_{\bm{r}}\delta\bm{B}(\bm{r})\cdot\bm{\sigma}(\bm{r}),\quad \bm{\sigma}(\bm{r})\equiv\bm{\sigma}n(\bm{r}).
\end{align}
It is convenient to define
\begin{align}
    [\mG_{u}]_{\bm{r}\bm{r}'}\equiv [\tilde{G}_{u}]_{\bm{r}\bm{r}'}e^{-i\delta A_{\bm{r}\bm{r}'}},
\end{align}
which allows us to rewrite the Dyson equation for $\tilde{G}_u$ in such a way that perturbation is manifestly gauge invariant. Indeed, by using $e^{i\delta A_{\bm{r}\bm{r}}}=1$ again, we obtain
\begin{align}
    \delta_{\bm{r}\bm{r}'}&=\sum_{\bm{r}''}[\tilde{G}_{u}^{-1}]_{\bm{r}\bm{r}''}[\tilde{G}_{u}]_{\bm{r}''\bm{r}'}e^{i\delta A_{\bm{r}'\bm{r}}}\label{eq:Dyson_mG}\\
    &=\sum_{\bm{r}''}[G_{u}^{-1}-\delta H_{\rm Z}]_{\bm{r}\bm{r}''}[\mG_{u}]_{\bm{r}''\bm{r}'}e^{i\delta\phi_{\bm{r}\bm{r}''\bm{r}'}},\notag
\end{align}
by following the procedure described in Ref.~\cite{Chen2011-dj}.
We defined the flux change
\begin{align}
    \delta\phi_{\bm{r}_1\bm{r}_2\bm{r}_3}
   &\equiv \delta A_{\bm{r}_1\bm{r}_2}+\delta A_{\bm{r}_2\bm{r}_3}+\delta A_{\bm{r}_3\bm{r}_1}\notag\\
    &\equiv\int_{\bar{\bm{x}}\in\triangle_{\bm{r}_1\bm{r}_2\bm{r}_3}}d\bm{S}\cdot\delta\bm{B}(\bar{\bm{x}}),\label{eq:Chen}
\end{align}
which threads the triangle $\triangle_{\bm{r}_1\bm{r}_2\bm{r}_3}$ formed by the three points $\bm{r}_1$, $\bm{r}_2$, and $\bm{r}_3$ with the normal direction defined appropriately.
Equation~\eqref{eq:Dyson_mG} indicates that the $\delta\bm{A}$ dependence of $\mG_{u}$ comes only through the magnetic field $\delta\bm{B}$ in the OBC geometry, and thus allows a perturbative  expansion in terms of the gauge-invariant small quantity.
[In PBCs, on the other hand, we have to care about the contributions from the paths winding the torus, though they are negligible due to the spatial decay of Matsubara Green's function for a large system.]
By writing $\mG_{u}=G_{u}+\delta\mG_{u}$, we obtain
\begin{align}
    [\delta\mG_{u}]_{\bm{r}\bm{r}'}&=-\sum_{\bm{r}_1\bm{r}_2}[G_{u}]_{\bm{r}\bm{r}_1}[G^{-1}_{u}]_{\bm{r}_1\bm{r}_2}[G_{u}]_{\bm{r}_2\bm{r}'}i\delta\phi_{\bm{r}_1\bm{r}_2\bm{r}'}\notag\\
    &\qquad\qquad+[G_u\delta H_{\rm Z}G_u]_{\bm{r}\bm{r}'}.\label{eq:dG}
\end{align}

Now we can derive the three-point formula for the local orbital magnetization, by combining Eqs.~\eqref{eq:cj}, \eqref{eq:deltaOmega_40a}, and \eqref{eq:dG}.
By denoting the trace over the internal degrees of freedom $\alpha$ as $\tr[\cdot]$, we obtain
\begin{align}
    \Tr[\tilde{G}_u]-\Tr[G_u]&=\sum_{\bm{r}}\tr[[\tilde{G}_u]_{\bm{r}\bm{r}}]-\Tr[G_u]\notag\\
    &=\sum_{\bm{r}}\tr[[\mG_u]_{\bm{r}\bm{r}}]-\Tr[G_u]\notag\\
    &=\Tr[\delta\mG_u].
\end{align}
Thus, we immediately obtain
\begin{align}
    \delta\Omega&=\frac{i}{\beta}\sum_{\omega_n}\int^0_{-\infty}du\sum_{\bm{r}_1\bm{r}_2\bm{r}_3}\tr\Bigl[[G_u^{-1}]_{\bm{r}_1\bm{r}_2}\notag\\
    &\qquad\qquad\qquad\cdot[G_u]_{\bm{r}_2\bm{r}_3}[G_u]_{\bm{r}_3\bm{r}_1}\Bigr]\delta\phi_{\bm{r}_1\bm{r}_2\bm{r}_3}\notag\\
    &\qquad -\frac{1}{\beta}\sum_{\omega_n}\int^0_{-\infty}du\,\Tr[G_u\delta H_{\rm Z}G_u].\label{eq:dOmega_temp}
\end{align}
Since $\delta\bm{B}(\bar{\bm{x}})=\nabla_{\bar{\bm{x}}}\times[\varepsilon\hat{x}_iW(\bm{x}-\bar{\bm{x}})]$, we can explicitly write the flux variation $\delta\phi$ as
\begin{align}
    \delta\phi_{\bm{r}_1\bm{r}_2\bm{r}_3}&=\epsilon_{ijk}\partial_{x_j}\left[\varepsilon\int_{\bar{\bm{x}}\in\triangle_{\bm{r}_1\bm{r}_2\bm{r}_3}} d\bm{S}\,W(\bm{x}-\bar{\bm{x}})\right]_k.\label{eq:dphi_temp}
\end{align}
Thus, we obtain the expression of the smeared local magnetization $\bar{\bm{m}}(\bm{x})$ by
\begin{subequations}\label{eq:cj_cm}
\begin{align}
    \braket{\bar{\bm{j}}(\bm{x})}&=\nabla\times\bar{\bm{m}}(\bm{x}),\\
    \bar{\bm{m}}(\bm{x})&=\int d^d\bar{{x}}\,W(\bm{x}-\bar{\bm{x}})\,\bm{m}_{\rm orb}(\bar{\bm{x}})\notag\\
&\qquad+\sum_{\bm{r}}W(\bm{x}-\bm{r})\bm{m}_{\rm spin}(\bm{r}),\label{eq:50b_SM}
\end{align}with
\begin{align}
    \bm{m}_{\rm spin}(\bm{r})\equiv\frac{1}{\beta}\sum_{\omega_n}e^{i\omega_n\eta}\Tr[\bm{\sigma}(\bm{r})G],
\end{align}
and
\begin{widetext}
\begin{align}
    {m}^i_{\rm orb}(\bm{x})&=\frac{i}{\beta}\sum_{\omega_n}e^{i\omega_n\eta}\int^0_{-\infty}du\sum_{\bm{r}_1\bm{r}_2\bm{r}_3}\tr\Bigl[n(\bm{r}_1)G_u^{-1}n(\bm{r}_2)G_un(\bm{r}_3)G_u\Bigr]
    {\chi}^i_{\bm{r}_1\bm{r}_2\bm{r}_3}(\bm{x}),\label{eq:formula_m_SM}
\end{align}
where $\chi^i_{\bm{r}_1\bm{r}_2\bm{r}_3}(\bm{x})$ is defined by, with $\bm{r}_{ij}\equiv\bm{r}_i-\bm{r}_j$,
\begin{align}
    \chi^i_{\bm{r}_1\bm{r}_2\bm{r}_3}(\bm{x})&\equiv\int_{\bar{\bm{x}}\in\triangle_{\bm{r}_1\bm{r}_2\bm{r}_3}}dS_i\,\delta(\bm{x}-\bar{\bm{x}})
     =\int_0^1ds\int_0^{1-s}ds'\,[\bm{r}_{31}\times\bm{r}_{21}]_i\delta(\bm{r}_1+s\bm{r}_{21}+s'\bm{r}_{31}-\bm{x})\label{eq:chi_defs_SM}
    \\&=[\bm{r}_{31}\times\bm{r}_{21}]_i\int_0^1ds_1 \int_0^1ds_2\int_0^1ds_3\,\delta(s_1+s_2+s_3-1)\delta(s_1\bm{r}_1+s_2\bm{r}_{2}+s_3\bm{r}_{3}-\bm{x}).\label{eq:temp60f}
\end{align}
\end{widetext}\end{subequations}
For example, in two dimensions, the function $\chi^i_{\bm{r}_1\bm{r}_2\bm{r}_3}(\bm{x})$ gives a unit vector either parallel or antiparallel to the out-of-plane direction when $\bm{x}$ is inside the triangle $\triangle_{\bm{r}_1\bm{r}_2\bm{r}_3}$, and vanishes otherwise.
In three dimensions, a delta-function component remains even after the surface integral, requiring $\bm{x}$ to be on the plane spanned by $\bm{r}_1,$ $\bm{r}_2$, and $\bm{r}_3$. Thus, practically,  $\chi^i_{\bm{r}_1\bm{r}_2\bm{r}_3}(\bm{x})$ should be calculated by replacing the delta function by e.g., Gaussian or Lorentzian with small spread, as is customary done for the evaluation of density of states. 
Equation~\eqref{eq:temp60f} is useful to see the symmetry of $\chi^i_{\bm{r}_1\bm{r}_2\bm{r}_3}(\bm{x})$: it behaves like the antisymmetric tensor $\epsilon_{ijk}$ against the permutations of $1,2,3$.

We take the Matsubara summation and the $u$ integral.
By using the spectral decomposition $H=\sum_m\epsilon_bp_b$, we can write
\begin{align}
    m^i_{\rm orb}(\bm{x})&=\sum_{\bm{r}_1\bm{r}_2\bm{r}_3}\chi^i_{\bm{r}_1\bm{r}_2\bm{r}_3}(\bm{x})m_{\bm{r}_1,\bm{r}_2,\bm{r}_3},
\end{align}
with
\begin{align}
    m_{\bm{r}_1,\bm{r}_2,\bm{r}_3}&\equiv i\sum_{mn}\Tr[n(\bm{r}_1)Hn(\bm{r}_2)p_mn(\bm{r}_3)p_n]\notag\\
    &\qquad\cdot\frac{1}{\beta}\sum_{\omega_n}e^{i\omega_n\eta}\int^0_{-\infty}du\,g_{n}g_m,
\end{align}
and $g_n=(i\omega_n+u-\epsilon_n)^{-1}$.
The Matsubara summation is performed as
\begin{align}
    \frac{1}{\beta}\sum_{\omega_n}e^{i\omega_n\eta}g_ng_m=\frac{f(\epsilon_n-u)-f(\epsilon_m-u)}{\epsilon_{nm}},
\end{align}
with the Fermi distribution $f(\epsilon)=(e^{\beta\epsilon}+1)^{-1}$.
Here and hereafter, the limit $\epsilon_n\to\epsilon_m$ is meant in the case of $n=m$.
The $u$ integral is performed by
\begin{align}
    -\int_{-\infty}^0 du\,f(\epsilon_n-u)=F(\epsilon_n),
\end{align}
with $F(\epsilon)=-\frac{1}{\beta}\Log(1+e^{-\beta\epsilon})$.
Thus, we obtain
\begin{align}
     \!\! m_{\bm{r}_1\bm{r}_2\bm{r}_3}=-i\sum_{mn}\frac{F_{mn}}{\epsilon_{mn}}\Tr[n(\bm{r}_1)Hn(\bm{r}_2)p_mn(\bm{r}_3)p_n],\label{eq:magkernel_SM}
\end{align}
as given in the main text.

\subsection{Three-point formula and the bond current}
Here we clarify the quantitative relationship between the bond current and the three-point formula.
We focus on a bond $\bm{r}_i\to\bm{r}_f$, and consider the bond current $\braket{J_{\bm{r}_i\to\bm{r}_f}}$ associated to this bond.
Let us define the coordinates $x_\parallel$ and $\bm{x}_\perp$ by
\begin{align}
    \bm{x}=\bm{r}_c+x_\parallel\hat{e}_\parallel+\bm{x}_\perp,\quad \bm{r}_c\equiv\frac{\bm{r}_f+\bm{r}_i}{2}.
\end{align}
Here, $\hat{e}_\parallel=\bm{r}_{\rm fi}/r_{\rm fi}$ is the unit vector along the bond, with $\bm{r}_{\rm fi}=\bm{r}_{\rm f}-\bm{r}_{\rm i}$ and $r_{\rm fi}=|\bm{r}_{\rm fi}|$.
The coordinate $-r_{\rm fi}/2\le x_\parallel\le r_{\rm fi}/2$ specifies the position along the bond, while $\bm{x}_\perp$ specifies the coordinates in the surface normal to the bond and satisfies $\bm{x}_\perp\cdot\hat{e}_\parallel=0$.
We focus on a value of $x_\parallel$ on the bond and integrate the smeared current density operator over a small area $S_{\rm fi}\equiv \{(x_\parallel,\bm{x}_\perp)|\,|\bm{x}_\perp|\le +0\}$.
When the smearing function $W(\bm{x})$ is chosen to be the delta function, we obtain
\begin{align}
    &\int_{S_{\rm fi}} d\bm{S}\cdot\bar{\bm{j}}(\bm{x})|_{W\to \delta}-(\text{spin contribution})\\
    &=-i\sum_{\bm{r}\bm{r}'}n(\bm{r})Hn(\bm{r}')\int^{\bm{r}}_{\bm{r}'}d\bar{\bm{x}}\cdot\hat{e}_\parallel\int_{S_{\rm fi}} d^{d-1}S\,\delta(\bm{x}-\bar{\bm{x}}).\notag
\end{align}
Here, the spin contribution on the first line specifies the contribution from the second term of Eq.~\eqref{eq:50b_SM}.
This integral is nonzero only when the line running from $\bm{r}'$ to $\bm{r}$ crosses the area $S_{\rm fi}$.
Thus, we obtain
\begin{align}
    (\text{Right hand side})&=-in(\bm{r}_{\rm f})Hn(\bm{r}_{\rm i})\int^{r_{\rm fi}/2}_{-r_{\rm fi}/2}d\bar{x}_\parallel \delta(x_\parallel-\bar{x}_\parallel)\notag\\
    &\qquad\qquad\qquad -(\bm{r}_{\rm i}\leftrightarrow \bm{r}_{\rm f})\notag\\
    &=J_{\bm{r}_{\rm i}\to\bm{r}_{\rm f}}.
\end{align}
On the other hand, we have
\begin{align}
    \braket{\bar{\bm{j}}(\bm{x})|_{W\to\delta}}-(\text{spin contribution})=\nabla_{\bm{x}}\times \bm{m}_{\rm orb}(\bm{x}).
\end{align}
We thus obtain
\begin{align}
    \braket{J_{\bm{r}_{\rm i}\to\bm{r}_{\rm f}}}&=\int_{S_{\rm fi}}d\bm{S}\cdot \nabla_{\bm{x}}\times\bm{m}_{\rm orb}(\bm{x})\notag\\
    &=\oint_{\partial S_{\rm fi}}d\bm{x}\cdot\bm{m}_{\rm orb}(\bm{x}).
\end{align}
Note also that $\oint_{\partial S}d\bm{x}\cdot m_{\rm orb}(\bm{x})=0$ if $\partial S$ does not encircle any of the bonds.

In two dimensions, it is convenient to define $\hat{e}_\perp\equiv\hat{z}\times\hat{e}_\parallel$ and write $\bm{x}_\perp=x_\perp\hat{e}_\perp$.
Then, we obtain
\begin{align}
    \braket{J_{\bm{r}_{\rm i}\to\bm{r}_{\rm f}}}=m^z_{\rm orb}(x_\parallel,x_\perp=+0)-m^z_{\rm orb}(x_\parallel,x_\perp=-0).
\end{align}
Note also that the local orbital magnetization remains constant as long as $\bm{x}$ does not cross a bond. Indeed, $\partial_im^z_{\rm orb}(\bm{x})=-\epsilon_{zij}\bar{j}_j(\bm{x})$ for $W(\bm{x})=\delta(\bm{x})$, and the right-hand side vanishes when $\bm{x}$ is off any of the bonds.

\section{Natural local marker}
Here we show the details of the derivation of the NLM.
The $\bm{x}$ dependence of the smeared magnetization is determined by
\begin{align}
    \bar{\chi}^i_{\bm{r}_1\bm{r}_2\bm{r}_3}(\bm{x})&\equiv\int_{\bar{\bm{x}}\in\triangle_{123}}dS_i\,W(\bm{x}-\bar{\bm{x}}).
\end{align}    
We rewrite $\bar{\chi}$ as $\bar{\chi}^i_{\bm{r}_1\bm{r}_2\bm{r}_3}(\bm{x})
    =\frac{1}{2}[\bm{r}_{12}\times\bm{r}_{32}]_i\chi_W$, with
\begin{align}    \chi_W&\equiv2\int_0^1ds\int_0^{1-s}ds'\,W\left(\bm{x}-[\bm{r}_1+s\bm{r}_{21}+s'\bm{r}_{31}]\right).
\end{align}
By noting that $\chi_W$ is invariant against cyclic permutation of $1,2,3$ in the integrand [see Eq.~\eqref{eq:temp60f}], we can approximate $\chi_W$ by
\begin{align}
    \chi_W&=\left[1- \frac{\bm{r}_{21}+\bm{r}_{31}}{3}\cdot\nabla_{\bm{x}}\right]W(\bm{x}-\bm{r}_1)+O(\nabla_{\bm{x}})^2,
\end{align}
or by its cyclic permutation.
We adopt the average of these three options, 
\begin{align}
    \chi_W\simeq\frac{1}{3}\sum_{a=1,2,3} \left[1- \frac{\bm{r}_{a+1,a}+\bm{r}_{a+2,a}}{3}\cdot\nabla_{\bm{x}}\right]W(\bm{x}-\bm{r}_a).\label{eq:temp77}
\end{align}
Here, the index $a$ is evaluated by modulo 3.

Plugging Eq.~\eqref{eq:temp77} into the three-point formula, we obtain an expansion of the form
\begin{align}
    \bar{m}^i_{\rm orb}(\bm{x})&=\bar{\mathcal{M}}_i(\bm{x})+\partial_{x_a}\bar{m}^{(2)}_{ia}(\bm{x})\notag\\
    &\qquad\qquad+\partial_{x_a}\partial_{x_b}\delta\bar{m}_{iab}(\bm{x}),
\end{align}
where all the higher-order contributions $m^{(3)}_{\rm orb}$, $m^{(4)}_{\rm orb}$, $\cdots$, are included in $\delta\bar{m}_{iab}(\bm{x})$.
We first focus on $\bar{\mathcal{M}}_i(\bm{x})$
and obtain
\begin{widetext}
    \begin{align}
        \bar{\mathcal{M}}_i(\bm{x})&=\frac{i}{6}\sum_{mn}\frac{F_{mn}}{\epsilon_{mn}}\sum_{\bm{r}_1\bm{r}_2\bm{r}_3}\sum_aW(\bm{x}-\bm{r}_a)[\bm{r}_{21}\times\bm{r}_{32}]_i\Tr[n(\bm{r}_1)Hn(\bm{r}_2)p_mn(\bm{r}_3)p_n]\notag\\
        &=\frac{i\epsilon_{ijk}}{6}\sum_{mn}\frac{F_{mn}}{\epsilon_{mn}}\sum_{\bm{r}_1\bm{r}_2\bm{r}_3}\sum_aW(\bm{x}-\bm{r}_a)\Tr[n(\bm{r}_1)[H,\hat{r}_j]n(\bm{r}_2)[p_m,\hat{r}_k]n(\bm{r}_3)p_n]\notag\\
        &=\frac{i\epsilon_{ijk}}{6}\sum_{mn}\frac{F_{mn}}{\epsilon_{mn}}\Bigl(\Tr[\bar{n}(\bm{x})[H,\hat{r}_j][p_m,\hat{r}_k]p_n]+\Tr[[H,\hat{r}_j]\bar{n}(\bm{x})[p_m,\hat{r}_k]p_n]+\Tr[[H,\hat{r}_j][p_m,\hat{r}_k]\bar{n}(\bm{x})p_n]\Bigr).
    \end{align}
    This gives the NLM in the main text,
    \begin{align}
    \mathcal{M}_i(\bm{r})&=\frac{\epsilon_{ijk}}{6}\sum_{mn}\frac{F_{mn}}{\epsilon_{mn}}\,\Tr\Bigl[n(\bm{r})\Bigl([\hat{r}_j,p_m]\,p_nJ_k+J_k[\hat{r}_j,p_m]\,p_n+p_nJ_k[\hat{r}_j,p_m]\Bigr)\Bigr]
    \label{eq:m1_SM}.
\end{align}

We next consider $\bar{m}^{(2)}_{ij}(\bm{x})$.
This is given by, with $\bm{r}_{21}\times\bm{r}_{32}=\bm{r}_{21}\times\bm{r}_{31}$,
\begin{align}
    \bar{m}^{(2)}_{ij}(\bm{x})&=-\frac{i\epsilon_{ijk}}{18}\sum_{mn}\frac{F_{mn}}{\epsilon_{mn}}\sum_{\bm{r}_1\bm{r}_2\bm{r}_3}\sum_a(\bm{r}_{a+1,a}+\bm{r}_{a+2,a})_lW(\bm{x}-\bm{r}_a)\tr[n(\bm{r}_1)Hn(\bm{r}_2)[\hat{r}_j,p_m]n(\bm{r}_3)[\hat{r}_k,p_n]]\notag\\
    &=-\frac{i\epsilon_{ijk}}{18}\sum_{mn}\frac{F_{mn}}{\epsilon_{mn}}\Bigl(\Tr[\bar{n}(\bm{x})[H,\hat{r}_l][\hat{r}_j,p_m][\hat{r}_k,p_n]]+\Tr[\bar{n}(\bm{x})H[\hat{r}_j,p_m][\hat{r}_l,[\hat{r}_k,p_n]]]\notag\\
    &\qquad\qquad+\Tr[H\bar{n}(\bm{x})[[\hat{r}_j,p_m],\hat{r}_l][\hat{r}_k,p_n]]+\Tr[[\hat{r}_l,H]\bar{n}(\bm{x})[\hat{r}_j,p_m][\hat{r}_k,p_n]]\notag\\
    &\qquad\qquad+\Tr[H[\hat{r}_j,p_m]\bar{n}(\bm{x})[[\hat{r}_k,p_n],\hat{r}_l]]+\Tr[H[\hat{r}_l,[\hat{r}_j,p_m]]\bar{n}(\bm{x})[\hat{r}_k,p_n]]\Bigr)\notag\\
    &=-\frac{i\epsilon_{ijk}}{18}\sum_{mn}\frac{F_{mn}}{\epsilon_{mn}}\Bigl(\Tr[[\bar{n}(\bm{x}),[H,\hat{r}_l]][\hat{r}_j,p_m][\hat{r}_k,p_n]]+\Tr[[\bar{n}(\bm{x}),H[\hat{r}_j,p_m]][\hat{r}_l,[\hat{r}_k,p_n]]]\notag\\
    &\qquad\qquad+\Tr[H[\bar{n}(\bm{x}),[[\hat{r}_j,p_m],\hat{r}_l]][\hat{r}_k,p_n]]\Bigr)\notag\\
    &=-\frac{i\epsilon_{ijk}}{18}\sum_{mn}\frac{F_{mn}}{\epsilon_{mn}}\Bigl(\Tr[[[\bar{n}(\bm{x}),H],\hat{r}_l][\hat{r}_j,p_m][\hat{r}_k,p_n]]+\Tr[[\bar{n}(\bm{x}),H][\hat{r}_j,p_m][\hat{r}_l,[\hat{r}_k,p_n]]]\notag\\
    &\qquad\qquad+\Tr[H[\hat{r}_j,[\bar{n}(\bm{x}),p_m]][\hat{r}_l,[\hat{r}_k,p_n]]]+\Tr[H[[\hat{r}_j,[\bar{n}(\bm{x}),p_m]],\hat{r}_l][\hat{r}_k,p_n]]\Bigr)\notag\\
     &=-\frac{i\epsilon_{ijk}}{18}\oint_C\frac{dz}{2\pi i}\Bigl(\Tr[[[\bar{n}(\bm{x}),H],\hat{r}_l][\hat{r}_j,G(z)][\hat{r}_k,O(z)]]+\Tr[[\bar{n}(\bm{x}),H][\hat{r}_j,G(z)][\hat{r}_l,[\hat{r}_k,O(z)]]]\notag\\
    &\qquad\qquad+\Tr[H[\hat{r}_j,[\bar{n}(\bm{x}),G(z)]][\hat{r}_l,[\hat{r}_k,O(z)]]]+\Tr[H[[\hat{r}_j,[\bar{n}(\bm{x}),G(z)]],\hat{r}_l][\hat{r}_k,O(z)]]\Bigr),
\end{align}
\end{widetext}
with $G(z)\equiv(z-H)^{-1}$, $O(z)\equiv(F(z)-F(H))/(z-H)$, and $[\hat{r}_j,\bar{n}(\bm{x})]=0$.
In this equation, the $\bm{x}$ dependence comes from either
\begin{align}
    [\bar{n}(\bm{x}),H]=-i\nabla_{\bm{x}}\cdot\bar{\bm{j}}(\bm{x}),
\end{align}
or
\begin{align}
    [\bar{n}(\bm{x}),G(z)]&=G(z)[\bar{n}(\bm{x}),H]G(z)\notag\\
    &=-i\nabla_{\bm{x}}\cdot[G(z)\bar{\bm{j}}(\bm{x})G(z)],
\end{align}
according to the continuity equation.
This means that $\bar{m}^{(2)}_{ia}(\bm{x})$ is a total derivative, and thus, we can write 
\begin{align}
    \bar{m}^{(2)}_{ia}(\bm{x})+\partial_{x_b}\delta\bar{m}_{iab}(\bm{x})=\partial_{x_b}\delta\bar{M}_{iab}(\bm{x}),
\end{align}
for some $\delta\bar{M}_{iab}(\bm{x})$.

    \section{Theree point formula in a bulk  reference point}
We here discuss 
how to evaluate the three-point formula for reference points in the bulk. While we consider lattice models, the continuum-model results are obtained by replacing the site summations with position integrals.
We focus on a bulk position $\bm{x}$ far from the boundaries, and start from
\begin{align}
    m^i_{\rm orb}(\bm{x})&=\sum_{\bm{r},\Delta\bm{r},\Delta\bm{r}'}\chi^i_{\bm{r}+\Delta\bm{r},\bm{r},\bm{r}+\Delta\bm{r}'}(\bm{x})m_{\bm{r};\Delta\bm{r},\Delta\bm{r}'},\label{eq:temp66}
\end{align}
with $m_{\bm{r};\Delta\bm{r},\Delta\bm{r}'}\equiv m_{\bm{r}+\Delta\bm{r},\bm{r},\bm{r}+\Delta\bm{r}'}$ and
\begin{align}
    m_{\bm{r};\Delta\bm{r},\Delta\bm{r}'}&=-i\oint_{C}\frac{dz}{2\pi i}\Tr\Bigl[n(\bm{r}+\Delta\bm{r})Hn(\bm{r})\notag\\
    &\cdot \left.\frac{1}{z-H}n(\bm{r}+\Delta\bm{r}')\frac{F(z)-F(H)}{z-H}\right].\label{eq:m_temp_67}
\end{align}
Here, the summation $\Delta\bm{r}$ and $\Delta\bm{r}'$ are taken over lattice sites around the site $\bm{r}$.
According to the locality of the magnetization kernel, only
the lattice sites where $|\bm{x}-\bm{r}|$, $|\Delta\bm{r}|$ and $|\Delta\bm{r}'|$ are much smaller than the system size contribute to the three-point formula, allowing us to limit the summation for $\Delta\bm{r}$ and $\Delta\bm{r}'$ to those with the absolute value much smaller than the system diameter, $|\Delta\bm{r}|,|\Delta\bm{r}'|\ll V^{1/d}$.
This locality also indicates that only the matrix elements of the Hamiltonian far away from the boundaries determine the magnetization, and therefore the results are insensitive to the boundary conditions.
Thus, we may consider extended systems instead of the finite-size flake with OBCs, or more practically, we may consider systems with PBCs with system size not smaller than that with OBCs.
Therefore, we can replace $H$ in Eq.~\eqref{eq:m_temp_67} with that with PBCs.

For clarity, we temporarily set the spatial origin of the finite-size flake to be near $\bm{x}$, and adopt the same coordinates for the PBCs under consideration.
Thus, every $\bm{x}$, $\bm{r}$, $\Delta\bm{r}$, and $\Delta\bm{r}'$ are much smaller than the system size.
The geometric factor is then given by
\begin{align}
    \chi^i_{\bm{r},\bm{r}+\Delta\bm{r},\bm{r}+\Delta\bm{r}'}(\bm{x})&=[\Delta\bm{r}\times\Delta\bm{r}']_i\int_0^1ds\int_0^{1-s}ds'\,\notag\\
    &\qquad\cdot\delta(\bm{r}+s\Delta\bm{r}+s'\Delta\bm{r}'-\bm{x})\notag\\
    &=[\Delta\bm{r}\times\Delta\bm{r}']_i\int_0^1ds\int_0^{1-s}ds'\,\notag\\
    &\qquad\cdot\delta'(\bm{r}+s\Delta\bm{r}+s'\Delta\bm{r}'-\bm{x}).\label{eq:temp68}
\end{align}
In the second line, we replaced the delta function with $\delta'$, which is the system-diameter-periodic extension of the delta function such that it is compatible with the PBCs under consideration.
This does not change the result, since the argument is much smaller than the system size in the present coordinate system.
The final expression, on the other hand, is free from the coordinate choices of the PBCs, and thus is applicable regardless of the coordinate systems.

Thus, we have seen that we can evaluate the three-point formula for a bulk reference point $\bm{x}$ of a finite-size flake with OBCs based on PBCs.
In particular, Eqs.~\eqref{eq:temp66} and \eqref{eq:m_temp_67} with $|\Delta\bm{r}|,|\Delta\bm{r}'|\ll V^{1/d}$, as well as the second line of Eq.~\eqref{eq:temp68}, gives the three-point magnetization field in systems with PBCs with a sufficiently large system size.

\subsection{Bulk orbital magnetization}
In the following, we show that the three-point formula at a bulk position $\bm{x}$ averaged over a unit cell coincides with the bulk orbital magnetization formula~\cite{Thonhauser2005-gi,Shi2007-rc}.
By writing $\bm{x}=\bm{R}_0+\delta\bm{x}$ with a lattice vector $\bm{R}_0$ and the position $\delta\bm{x}$ within a unit cell, we are thus interested in
\begin{align}
    {m}^i_{\rm bulk}(\bm{R}_0)\equiv\frac{1}{V_{\rm UC}}\int_{\text{unit cell}}d^d[\delta x]\,m^i_{\rm orb}(\bm{R}_0+\delta\bm{x}),\label{eq:temp_morb_bulk}
\end{align}
where $V_{\rm UC}$ is the unit-cell volume.
Assuming the system without disorder, the Hamiltonian $H$ have the crystal translational symmetry. Thus, the magnetization kernel is invariant under the shift of an arbitrary translation vector $\bm{R}$, to obtain \begin{align}
    m_{\bm{r},\Delta\bm{r},\Delta\bm{r}'}&=m_{\bm{r}+\bm{R},\Delta\bm{r},\Delta\bm{r}'}\notag\\
    &=\frac{1}{N_{\rm UC}}\sum_{\bm{R}'}m_{\bm{r}+\bm{R}',\Delta\bm{r},\Delta\bm{r}'},
\end{align}
with the number of unit cells $N_{\rm UC}$.
This erases the $\bm{R}_0$ dependence of eq.~\eqref{eq:temp_morb_bulk},
\begin{align}
    &m_{\rm bulk}^i=\frac{1}{V_{\rm UC}}\int_{\text{unit cell}}d^d[\delta x]\,m^i_{\rm orb}(\bm{R}_0+\delta\bm{x})\notag\\
    &=\frac{1}{V_{\rm UC}}\sum_{\bm{r},\Delta\bm{r},\Delta\bm{r}'}m_{\bm{r};\Delta\bm{r},\Delta\bm{r}'}[\Delta\bm{r}\times\Delta\bm{r}']_i\int_0^1d^ds\int_0^{1-s}d^ds'\notag\\
    &\quad\cdot\frac{1}{N_{\rm UC}}\sum_{\bm{R}'}\int_{\text{unit cell}}d^d[\delta x]\delta'(\bm{r}-\bm{R}'+s\Delta\bm{r}+s'\Delta\bm{r}'-\delta\bm{x})\notag\\
    &=\frac{1}{V}\sum_{\bm{r},\Delta\bm{r},\Delta\bm{r}'}m_{\bm{r};\Delta\bm{r},\Delta\bm{r}'}[\Delta\bm{r}\times\Delta\bm{r}']_i\int_0^1d^ds\int_0^{1-s}d^ds'\,1\notag\\
    &=\frac{1}{2V}\sum_{\bm{r},\Delta\bm{r},\Delta\bm{r}'}m_{\bm{r};\Delta\bm{r},\Delta\bm{r}'}[\Delta\bm{r}\times\Delta\bm{r}']_i,
\end{align}
with the system volume $V=V_{\rm UC}N_{\rm UC}$.
Here, the site difference appears in the form
\begin{align}
\Delta\bm{r}\braket{\bm{r}+\Delta\bm{r}\alpha|H|\bm{r}\beta},\quad
\Delta\bm{r}'\braket{\bm{r}\alpha|G(z)|\bm{r}+\Delta\bm{r}'\beta},
\end{align}
and can be rewritten by using the derivatives in terms of the spatially-uniform vector potential $\partial_{A_j}$,
\begin{subequations}\begin{align}
    i\Delta\bm{r}\braket{\bm{r}+\Delta\bm{r}\alpha|H|\bm{r}\beta}&=\braket{\bm{r}+\Delta\bm{r}\alpha|\partial_{\bm{A}}H|\bm{r}\beta},\\
\Delta\bm{r}'\braket{\bm{r}\alpha|G(z)|\bm{r}+\Delta\bm{r}'\beta}&\simeq i\braket{\bm{r}\alpha|\partial_{\bm{A}}G(z)|\bm{r}+\Delta\bm{r}'\beta},
\end{align}\label{eq:GFPBCOBC}\end{subequations}
the latter of which follows 
since the neglected contributions are  exponentially small with respect to the system size.

After taking the $z$ integral, or after taking the Matsubara summation based on Eq.~\eqref{eq:formula_m_SM} as illustrated in the next section for NLM, we obtain 
\begin{align}
    m^i_{\rm bulk}&=-\frac{i\epsilon_{ijk}}{2}\sum_{mn}\int\frac{d^dk}{(2\pi)^d}\frac{F(\epsilon_{\bm{k}m})-F(\epsilon_{\bm{k}n})}{\epsilon_{\bm{k}m}-\epsilon_{\bm{k}n}}\notag\\
    &\qquad\cdot\tr[\partial_{k_j}H_{\bm{k}}\partial_{k_k}P_{\bm{k}m}P_{\bm{k}n}].\label{eq:Mz_PBC_68}
\end{align}
Here, we re-labeled the eigenstates as $m\to (\bm{k},m)$, where $\bm{k}$ is the crystal momentum and $m$ is redefined to be the band index in the following part of this section.
We defined the Bloch Hamiltonian $H_{\bm{k}}\equiv e^{-i\bm{k}\cdot\hat{\bm{r}}}He^{i\bm{k}\cdot\hat{\bm{r}}}$ and the Bloch state $\ket{\psi_{\bm{k}m}}=e^{i\bm{k}\cdot\hat{\bm{r}}}\ket{u_{\bm{k}m}}$.
The projector is then given by $p_{\bm{k}m}=e^{i\bm{k}\cdot\hat{\bm{r}}}P_{\bm{k}m}e^{-i\bm{k}\cdot\hat{\bm{r}}}$, with $P_{\bm{k}m}\equiv\ket{u_{\bm{k}m}}\bra{u_{\bm{k}m}}$.
In the presence of degeneracy, $P_{\bm{k}m}$ should include all the degenerate states with energy $\epsilon_{\bm{k}m}$. In the following, however, we assume that the bands are not degenerate for simplicity.
The trace $\tr[\cdot]$ is taken over all the energy bands, or equivalently, 
the sublattices and internal degrees of freedom.

By using $\tr[H_{\bm{k}}\partial_{k_k}P_{\bm{k}m}P_{\bm{k}n}]=0$, we obtain
\begin{align}
    m^i_{\rm bulk}
    &=-\frac{i\epsilon_{ijk}}{2}\sum_{mn}\int\frac{d^dk}{(2\pi)^d}\frac{F(\epsilon_{\bm{k}m})-F(\epsilon_{\bm{k}n})}{\epsilon_{\bm{k}m}-\epsilon_{\bm{k}n}}\notag\\
    &\qquad\cdot\tr[H_{\bm{k}}\partial_{k_j}P_{\bm{k}m}\partial_{k_k}P_{\bm{k}n}].
\end{align}
Note that we have the identity
\begin{align}
    &\frac{i\epsilon_{ijk}}{2}\tr[H_{\bm{k}}\partial_{k_j}P_{\bm{k}m}\partial_{k_k}P_{\bm{k}n}]\notag\\&=\delta_{mn}m_{\bm{k}n}^i -\frac{1}{2}(\epsilon_{\bm{k}m}-\epsilon_{\bm{k}n})\omega_{mn}^i(\bm{k}),
\end{align}
with 
\begin{subequations}\begin{gather}
   m^i_{\bm{k}n}\equiv\frac{i\epsilon_{ijk}}{2}\braket{\partial_{k_j}u_{\bm{k}n}|\epsilon_{\bm{k}n}-H_{\bm{k}}|\partial_{k_k}u_{\bm{k}n}},\\
   \omega_{mn}^i(\bm{k})\equiv i\epsilon_{ijk}A_{mn}^j(\bm{k})A_{nm}^k(\bm{k}),\\    A_{nm}^j(\bm{k})=i\braket{u_{\bm{k}n}|\partial_{k_j}u_{\bm{k}m}}.   
\end{gather}\end{subequations}
By using $F_{mn}/\epsilon_{mn}\to f(\epsilon_m)$ for $m=n$ and introducing the Berry curvature
\begin{align}
    \Omega_m^i(\bm{k})\equiv\sum_{n}\omega_{mn}^i(\bm{k}),
\end{align}
we reproduce the bulk orbital magnetization
\begin{align}
    \!\!\!\!\!\!m^i_{\rm orb}&=-\sum_m\int\frac{d^dk}{(2\pi)^d}f(\epsilon_{\bm{k}m})m_{\bm{k}m}^i-F(\epsilon_{\bm{k}m})\Omega_m^i(\bm{k}),
\end{align}
by setting $e/\hbar\to-1$ in the formula given in Ref.~\cite{Shi2007-rc}.

\section{PBC formula for the NLM\label{sec:PBC}}
When the lattice site $\bm{r}$ is away from the boundaries, we can express the NLM by using the quantities in PBCs.
Here we discuss the Bloch-state formulas of NLM for a bulk site $\bm{r}=\bm{R}+\delta\bm{r}$.
By using the Green's function, NLM is given by
\begin{align}
    \mathcal{M}_j(\bm{r})&=\frac{T\epsilon_{jkl}}{6}\sum_{\omega_n}e^{i\omega_n\eta}\int^0_{-\infty} du\,\Tr[n(\bm{r})([\hat{r}_k,G_u]J_lG_u\notag\\
    &\qquad+J_lG_u[\hat{r}_k,G_u]+G_u[\hat{r}_k,G_u]J_l)].
\end{align}
As discussed near Eq.~\eqref{eq:GFPBCOBC}, Green's function can be replaced with that in PBCs when $\bm{r}$ is a bulk lattice site, and the commutator of the position operator and the Green's function can be replaced with the vector-potential derivative.
In the Bloch representation, Green's function can be expressed as
\begin{align}
    G_u=\sum_{\bm{k}m}\frac{1}{i\omega_n+u-\epsilon_{\bm{k}-\bm{A},m}}e^{i\bm{k}\cdot\hat{\bm{r}}}P_{\bm{k}-\bm{A},m}e^{-i\bm{k}\cdot\hat{\bm{r}}},
\end{align}
with a spatially-uniform vector potential $\bm{A}$.
By using this,
wave-number derivative appears instead of the vector-potential derivatives in the Bloch representation, 
and thus we obtain
\begin{align}
    &\mathcal{M}_j(\bm{r})=\frac{iT\epsilon_{jkl}V_{\rm UC}}{6}\sum_{\omega_n}e^{i\omega_n\eta}\int_{-\infty}^0 du\int\frac{d^dk}{(2\pi)^d}\\
    &\cdot\tr[p_{\delta\bm{r}}(\partial_{k_k}G_{\bm{k}}J_{\bm{k}l}G_{\bm{k}}+J_{\bm{k}l}G_{\bm{k}}\partial_{k_k}G_{\bm{k}}+G_{\bm{k}}\partial_{k_k}G_{\bm{k}}J_{\bm{k}l})].\notag
\end{align}
Here, according to the translational symmetry, we replaced $n(\bm{r})$ with $p_{\delta\bm{r}}/N_{\rm UC}$, where $p_{\delta\bm{r}}$ is the projection operator to the sublattice to which the site $\bm{r}$ belongs.
Note that Eq.~\eqref{eq:Mz_PBC_68} can be obtained by summing over $p_{\delta\bm{r}}$ and thus by replacing $p_{\delta\bm{r}}$ with unity in this equation, after divided by the unit-cell volume $V_{\rm UC}$ to obtain the magnetization rather than magnetic moment.
After performing the Matsubara summation and the $u$ integral, we obtain
\begin{align}
    &\mathcal{M}_j(\bm{r})=\frac{i\epsilon_{jkl}V_{\rm UC}}{6}\int\frac{d^dk}{(2\pi)^d}\sum_{mn}\Bigl\{-\frac{F_{mn}}{\epsilon_{mn}}\notag\\
   & \cdot\tr[p_{\delta\bm{r}}(\partial_{k_k}P_m\partial_{k_l}HP_n+\partial_{k_l}HP_n\partial_{k_k}P_m+P_n\partial_{k_k}P_m\partial_{k_l}H)]\notag\\
    &\quad+\partial_{k_k}\epsilon_m\left(-\frac{f_m}{\epsilon_{mn}}+\frac{F_{mn}}{\epsilon_{mn}^2}\right)\tr[p_{\delta\bm{r}}(P_m\partial_{k_l}HP_n\notag\\
    &\qquad\qquad+\partial_{k_l}HP_nP_m+P_nP_m\partial_{k_l}H)]\Bigr\}.\label{eq:91_SM}
\end{align}
Here the wave-number dependence is implicit.

To simplify the expression, it is convenient to take the real part of both sides, which does not change the result since $\mathcal{M}_j(\bm{r})$ is real. This forces the last two terms of Eq.~\eqref{eq:91_SM} to cancel with each other.
We can also use
\begin{align}
    \partial_{k_l}HP_n&=\partial_{k_l}\epsilon_nP_n+(\epsilon_n-H)\partial_{k_l}P_n,
\end{align}
and thus $P_m\partial_{k_l}HP_n=\delta_{nm}\partial_{k_l}\epsilon_mP_m-\epsilon_{mn}P_m\partial_{k_l}P_n$.
After some algebra, we obtain
\begin{align}
   \mathcal{M}_j(\bm{r})&=\frac{\epsilon_{jkl}V_{\rm UC}}{6}\Im\int\frac{d^dk}{(2\pi)^d}\Bigl\{\sum_mf_m\tr[p_{\delta\bm{r}}\partial_{k_k}P_m\partial_{k_l}H]\notag\\
   &\qquad+\sum_{mn}\frac{F_{mn}}{\epsilon_{mn}}\tr[p_{\delta\bm{r}}(\partial_{k_k}P_m(\epsilon_n-H)\partial_{k_l}P_n\notag\\
   &\qquad+2(\epsilon_n-H)\partial_{k_l}P_n\partial_{k_k}P_m)]\Bigr\}.
\end{align}
This can also be expressed as
\begin{align}
    \mathcal{M}^i(\delta\bm{r})&=\frac{V_{\rm UC}}{6}\int\frac{d^dk}{(2\pi)^d}\Bigl(\sum_{mn}O_{mn}(\bm{k})S_{nm}^{i,\delta\bm{r}}(\bm{k})\\
    &\quad{+}\frac{\epsilon_{ijk}}{2}\Im\sum_mf_{\bm{k}m}\Tr[p_{\delta\bm{r}}[\partial_{k_j}P_{\bm{k}m},\partial_{k_k}H_{\bm{k}}]]\Bigr).\notag
\end{align}
Here, 
$O_{mn}(\bm{k})\equiv (F(\epsilon_{\bm{k}m})-F(\epsilon_{\bm{k}n}))/(\epsilon_{\bm{k}m}-\epsilon_{\bm{k}n})$, $f_{\bm{k}m}\equiv f(\epsilon_{\bm{k}m})$, and
\begin{align}
    S_{nm}^{i,\delta\bm{r}}(\bm{k})&\equiv\epsilon_{ijk}\Im\Tr[p_{\delta\bm{r}}(2\partial_{k_j}P_{\bm{k}m}\partial_{k_k}P_{\bm{k}n}(H_{\bm{k}}-\bar{\epsilon}_{\bm{k},mn})\notag\\
    &\quad+\partial_{k_k}P_{\bm{k}n}(H_{\bm{k}}-\bar{\epsilon}_{\bm{k},mn})\partial_{k_j}P_{\bm{k}m})],
\end{align}
with $\bar{\epsilon}_{\bm{k},mn}\equiv(\epsilon_{\bm{k}m}+\epsilon_{\bm{k}n})/2$.

We can further simplify the formula for two-band models such as the Haldane model.
We consider the Bloch Hamiltonian $H_{\bm{k}}=d_{0\bm{k}}+\bm{d}_{\bm{k}}\cdot\bm{\sigma}$, where the matrix structure corresponds to the sublattice degrees of freedom and the projector to each sublattice $s=\pm1$ is given by $p_s=(1+s\sigma_z)/2$.
Then, we can show that the Bloch-state formula reduces to
\begin{align}
    \mathcal{M}^i_s&=\frac{V_{\rm UC}}{2}\int\frac{d^dk}{(2\pi)^d}\Bigl(\mathcal{F}({\bm{k}})d_{\bm{k}}\Omega^i_{-}(\bm{k})+\frac{s\delta f_{\bm{k}}}{3}X^i(\bm{k})\Bigr),\label{eq:LM_bulk_formula}
\end{align}
with $d_{\bm{k}}\equiv|\bm{d}_{\bm{k}}|$, $\mathcal{F}(\bm{k})\equiv f_{\bm{k}+}+f_{\bm{k}-}-2O_{+-}(\bm{k})$, $\delta f_{\bm{k}}\equiv f_{\bm{k}+}-f_{\bm{k}-}$. We also defined the Berry curvature
$\Omega_-^i(\bm{k})\equiv\frac{\epsilon_{ijk}}{4d_{\bm{k}}^3}\bm{d}_{\bm{k}}\cdot\partial_j\bm{d}_{\bm{k}}\times\partial_j\bm{d}_{\bm{k}},$
and $X^i(\bm{k})\equiv\frac{\epsilon_{ijk}}{4d_{\bm{k}}}\hat{z}\cdot\partial_{j}\bm{d}_{\bm{k}}\times\partial_k\bm{d}_{\bm{k}}$.
The first term of Eq.~\eqref{eq:LM_bulk_formula} contributes to the bulk orbital magnetization $\sum_s\mathcal{M}_s^i/V_{\rm UC}$, while the second term describes the sublattice-staggered contribution.
Since $X^i(\bm{k})=d_{z\bm{k}}\Omega_-^i(\bm{k})$, Berry curvature is required to obtain a nonzero NLM at bulk lattice sites of two-band models.

\section{Magnetic quadrupole moment}
Here we discuss MQM in OBCs.
This is defined by
\begin{align}
    M_{\rm orb}^{ij}=\frac{1}{3}\int d^dx\,x_i[\bm{x}\times\braket{\bar{\bm{j}}(\bm{x})}]_j.\label{eq:MQM_def_SM}
\end{align}
The spatial integration is over $\bm{R}^d$, including the space outside of the system, where $\braket{\bar{\bm{j}}(\bm{x})}$ tends to vanish as going away from the system with the length scale of the smearing function $W$.
By using $\braket{\bar{j}_i(\bm{x})}=\epsilon_{ijk}\partial_{x_j}\bar{m}_k(\bm{x})$ and $\bar{m}_k(\bm{x})\to0$ for $|\bm{x}|\to\infty$, we obtain an alternative expression
\begin{align}
    M_{\rm orb}^{ij}&=\int d^dx\,x_i\bar{m}_j(\bm{x})-\frac{\delta_{ij}}{3}x_k\bar{m}_k(\bm{x}),\label{eq:MQM_mbased_SM}
\end{align}
both for $d=2$ and $3$.
Here, $\bar{m}_j(\bm{x})$ is the field given by smearing the three-point formula.

We can also express MQM based on the NLM.
Indeed, by substituting $\bar{m}_j(\bm{x})=\bar{\mathcal{M}}_j(\bm{x})+\partial_{x_a}\partial_{x_b}\delta\mathcal{M}(\bm{x})$, we obtain
\begin{align}
    M_{\rm orb}^{ij}&=\int d^dx\,x_i\bar{\mathcal{M}}_j(\bm{x})-\frac{\delta_{ij}}{3}x_k\bar{\mathcal{M}}_k(\bm{x}).
\end{align}
This is equivalent to 
\begin{align}
    M_{\rm orb}^{ij}&=\sum_{\bm{r}}r_i\mathcal{M}_j(\bm{r})-\frac{\delta_{ij}}{3}r_k{\mathcal{M}}_k(\bm{r}),\label{eq:Qij_SM}
\end{align}
by assuming 
either the thermodynamic limit or $W(\bm{x})=W(-\bm{x})$.
This gives a microscopic formula for the orbital MQM based on the NLM.

We can also have a current-operator-based quantum-mechanical formula for the orbital MQM that is expressed as the equilibrium expectation value of a quadrupole operator. By using the smeared current-density operator with Eq.~\eqref{eq:MQM_def_SM}, we can write
\begin{align}
    M_{\rm orb}^{ij}&=-\frac{i\epsilon_{jkl}}{3}\sum_{\bm{r}\bm{r}'}\braket{n(\bm{r})Hn(\bm{r}')}\int^{\bm{r}}_{\bm{r}'}d\bar{x}_l\notag\\
    &\qquad\qquad\cdot\int d^dx\,x_ix_k\,W(\bm{x}-\bar{\bm{x}}).
\end{align}
Assuming that $W(\bm{x})$ is isotropic, we obtain
\begin{align}
    \int d^dx\,x_ix_kW(\bm{x}-\bar{\bm{x}})=\bar{x}_i\bar{x}_k+\sigma_W\delta_{ik},
\end{align}
with a constant $\sigma_W$.
The contribution of the second term vanishes since it is proportional to the total current in equilibrium.
The contribution from the first term appears via
\begin{align}
    \int^{\bm{r}}_{\bm{r}'}d\bar{x}_l\,\bar{x}_i\bar{x}_k&=(r_l-r'_l)\Bigl[\frac{r_i+r'_i}{2}\frac{r_k+r'_k}{2}\notag\\
    &\quad+\frac{1}{12}(r_i-r'_i)(r_k-r_k')\Bigr],
\end{align}
whose second term does not contribute to $M_{\rm orb}^{ij}$ after antisymmetrization by $\epsilon_{jkl}$.
Thus, we obtain
\begin{align}
    M_{\rm orb}^{ij}&=-\frac{i\epsilon_{jkl}}{3}\sum_{\bm{r}\bm{r}'}\braket{n(\bm{r})Hn(\bm{r}')}(r_l-r_l')\frac{r_i+r_i'}{2}\frac{r_k+r_k'}{2}\notag\\
    &=\frac{\epsilon_{jkl}}{3}\sum_{\bm{r}\bm{r}'}\braket{n(\bm{r})i[H,\hat{r}_l]n(\bm{r}')}\frac{r_i+r_i'}{2}\frac{r_k+r_k'}{2}\notag\\
    &=\frac{1}{12}\braket{\{\hat{r}_i,\epsilon_{jkl}\{\hat{r}_k,J_l\}\}}.\label{eq:Mij_SM_ave}
\end{align}

From the procedure in the above, we know that Eqs.~\eqref{eq:Qij_SM} and \eqref{eq:Mij_SM_ave} coincide with each other, i.e.,
\begin{align}
    \sum_{\bm{r}}{r}_i\mathcal{M}_j(\bm{r})-\frac{\delta_{ij}}{3}r_k\mathcal{M}_k(\bm{r})&=\frac{1}{12}\braket{\{\hat{r}_i,\epsilon_{jkl}\{\hat{r}_k,J_l\}\}}.\label{eq:identity_MQM}
\end{align}
This is an exact identity even for small systems, as understood from their independence of $W(\bm{x})$. 
While coarse-graining procedure, which has been considered in deriving the NLM, looks meaningless  for small systems e.g., with diameter comparable to lattice constant, it is still possible to formally consider the coarse graining by $W(\bm{x})$ whose length scale is much larger than the system size. Then, the expansion of $W$ to obtain the NLM remains valid, and the equality of Eqs.~\eqref{eq:Qij_SM} and \eqref{eq:Mij_SM_ave} follows in the same way, by noting that the length scale of $W$ is not important as long as $W$ is chosen to be isotropic. 

It is also possible to directly show the identity~\eqref{eq:identity_MQM}.
We start from the expression based on Green's function,
\begin{align}
    \mathcal{M}_j(\bm{r})&=\frac{T\epsilon_{jkl}}{6}\sum_{\omega_n}\int^0_{-\infty} du\,\Tr[n(\bm{r})([r_k,G_u]J_lG_u\notag\\
    &\qquad+J_lG_u[r_k,G_u]+G_u[r_k,G_u]J_l)].
\end{align}
We also have
\begin{align}
    &\frac{1}{12}\braket{\{\hat{r}_i,\epsilon_{jkl}\{\hat{r}_k,J_l\}\}}\notag\\&=-\frac{T\epsilon_{jkl}}{12}\sum_{\omega_n}\int^0_{-\infty }du\Tr[\{\hat{r}_i,\epsilon_{jkl}\{\hat{r}_k,J_l\}\}G_u^2]\notag\\
    &=-\frac{T\epsilon_{jkl}}{6}\sum_{\omega_n}\int^0_{-\infty }du\Tr[\{\hat{r}_i,\epsilon_{jkl}\hat{r}_kJ_l\}G_u^2],
\end{align}
by using $\epsilon_{jkl}[\hat{r}_k,J_l]=0$. We obtain
\begin{align}
    &\sum_{\bm{r}}r_i\mathcal{M}_j(\bm{r})-\frac{1}{12}\braket{\{\hat{r}_i,\epsilon_{jkl}\{\hat{r}_k,J_l\}\}}\notag\\
    &=\frac{T}{6}\sum_{\omega}\int^0_{-\infty} du\,\Delta_{ij}+\epsilon_{jkl}\Tr[\hat{r}_i\hat{r}_kG_uJ_lG_u],\label{eq:target}
\end{align}
with
\begin{align}
\Delta_{ij}&\equiv i\epsilon_{jkl}\Tr[-\hat{r}_iG_u\hat{r}_k[\hat{r}_l,G_u^{-1}]G_u\notag\\
& +\hat{r}_i[\hat{r}_l,G_u^{-1}]G_u\hat{r}_kG_u+\hat{r}_iG_u\hat{r}_kG_u[\hat{r}_l,G_u^{-1}]],
\end{align}
by using $J_l=i[\hat{r}_l,G_u^{-1}]$.
The term other than $\Delta_{ij}$ vanishes since $\Tr[\hat{r}_i\hat{r}_kG_uJ_lG_u]=i\Tr[\hat{r}_i\hat{r}_k[G_u,\hat{r}_l]]=0$.
Furthermore, we can show that $\Delta_{ij}$ is proportional to $\delta_{ij}$ as follows:
\begin{align}
    \Delta_{ij}&=i\epsilon_{jkl}\Tr[r_iG_ur_kG_u^{-1}r_lG_u\notag\\
    &\qquad\qquad-r_iG^{-1}_ur_lG_ur_kG_u+r_iG_ur_kG_ur_lG_u^{-1}]\notag\\
    &=\frac{i\epsilon_{jkl}}{2}\Tr[G_ur_iG_u(r_kG_u^{-1}r_l-r_lG_u^{-1}r_k)\notag\\
    &\quad \qquad\qquad+G_ur_kG_u(r_lG_u^{-1}r_i-r_iG_u^{-1}r_l)\notag\\
    &\quad \qquad\qquad+G_ur_lG_u(r_iG_u^{-1}r_k-r_kG_u^{-1}r_i)].
\end{align}
The quantities in the trace is totally antisymmetric with respect to $i,k,l$, and thus is proportional to $\epsilon_{ikl}$, leading to $\Delta_{ij}\propto\epsilon_{jkl}\epsilon_{ikl}=2\delta_{ij}$.
It follows that the right-hand side of Eq.~\eqref{eq:target} is proportional to $\delta_{ij}$, and is 
identified as the trace part of $\sum_{\bm{r}}r_i\mathcal{M}_j(\bm{r})$, since $\frac{1}{12}\braket{\{\hat{r}_i,\epsilon_{jkl}\{\hat{r}_k,J_l\}\}}$ is traceless.
Thus, we directly reproduced the formula Eq.~\eqref{eq:identity_MQM}.

Before proceeding to the next section, we derive the MQM formula based on the three-point formula.
We can take the limit $W(\bm{x})\to \delta(\bm{x})$ in Eq.~\eqref{eq:MQM_mbased_SM}, since it is independent of the smearing function $W$ as long as it is isotropic as ensured by Eq.~\eqref{eq:Mij_SM_ave}.
We thus obtain with the three-point formula $m_j(\bm{x})$,
\begin{align}
    M_{\rm orb}^{ij}=\int d^dx\,x_i{m}_j(\bm{x})-\frac{\delta_{ij}}{3}x_k{{m}}_k(\bm{x}).
\end{align}
Similar magnetization-based formulas can also be obtained for higher-order multipole moments.
By performing the $x$ integration, we obtain
\begin{align}
    M_{\rm orb}^{ij}&=\sum_{\bm{r}_1\bm{r}_2\bm{r}_3}\frac{r_1^i+r_2^i+r_3^i}{6}\epsilon_{jkl}r^k_{31}r^l_{21}m_{\bm{r}_1\bm{r}_2\bm{r}_3}-(\text{trace part})\notag\\
    &=\sum_{\bm{r}}r_i\mathcal{M}_j(\bm{r})-\frac{\delta_{ij}}{3}r_k\mathcal{M}_k(\bm{r}),
\end{align}
which coincides with the formula obtained from the NLM.

\section{Details of the model}
Here we show the details of the model calculations.
We adopt the Haldane model~\cite{Haldane1988-og},
\begin{subequations}\begin{align}
    H=H_{\rm lc}+H_{\rm NN}+H_{\rm NNN},
\end{align}
with
\begin{align}
    H_{\rm lc}\equiv \sum_{\bm{R}}(v_{\rm SL}-\mu)\ket{\bm{R}_A}\bra{\bm{R}_A}+(-v_{\rm SL}-\mu)\ket{\bm{R}_B}\bra{\bm{R}_B},
\end{align}
and
\begin{align}
    H_{\rm NN}&\equiv -t_1\sum_{\bm{R};i=1,2,3}\ket{\bm{R}_A+\bm{\delta}_i}\bra{\bm{R}_A}+\text{h.c.},\\
    H_{\rm NNN}&\equiv-t_2\sum_{\bm{R};i=1,2,3}e^{i\phi}\ket{\bm{R}_A+\tilde{\bm{a}_i}}\bra{\bm{R}_A}\notag\\
    &\quad\quad+e^{-i\phi}\ket{\bm{R}_B+\tilde{\bm{a}}_i}\bra{\bm{R}_B}+\text{h.c.}.
\end{align}
\end{subequations}
Here, we set the unit-cell coordinates
\begin{subequations}\begin{gather}
    \bm{R}\in \mathbb{Z}\bm{a}_1+\mathbb{Z}\bm{a}_2,\\
    \bm{a}_1=\sqrt{3}a(1,0),\quad \bm{a}_2=\sqrt{3}a(1/2,\sqrt{3}/2),
\end{gather}
and the sublattice positions
\begin{align}
    \delta\bm{r}_A\equiv a(\sqrt{3}/2,-1/2),\quad \delta\bm{r}_B=a(\sqrt{3}/2,1/2),
\end{align}
to define $\bm{R}_A\equiv \bm{R}+\delta\bm{r}_A$ and $\bm{R}_B\equiv \bm{R}+\delta\bm{r}_B$.
We also define the nearest-neighbor bonds
\begin{align}
    \bm{\delta}_1=a(0,1),\quad \bm{\delta}_2=a(-\sqrt{3}/2,-1/2),
\end{align}
\end{subequations}
and $\bm{\delta}_3=-\bm{\delta}_1-\bm{\delta}_2$, as well as $\tilde{\bm{a}}_1=\bm{a}_1$, $\tilde{\bm{a}}_2=-\bm{a}_1+\bm{a}_2$, and $\tilde{\bm{a}}_3=-\tilde{\bm{a}}_1-\tilde{\bm{a}}_2$.
In the Hamiltonian, each component is kept only when all the ket and bra states $\ket{\bm{r}}$ and $\bra{\bm{r}}$ appearing in the component specify the lattice sites included in the system with OBCs as specified below.

We consider finite-size OBC flakes with zigzag and armchair edges in the $x$ and $y$ directions, respectively.
They are constructed as follows.
We put a home hexagon with its centroid at the origin $\bm{x}=0$.
We place $N_x$ hexagons on the right and left of the home hexagon to form a row of $2N_x+1$ hexagons whose centroids are on the line $y=0$.
We put a row of $2N_x+2$ hexagons on the top and bottom sides of this row, and then rows of $2N_x+1$ and $2N_x+2$ hexagons, repeatedly, until there are $N_y$ rows on the top and bottom sides of the central row.
For example, Fig.1(a) in the main text corresponds to $N_x=N_y=1$.
The number of lattice sites with the size-$(N_x,N_y)$ flake is
\begin{align}
    N_{\rm site}=\begin{cases}2(4N_x+5)(N_y+1)&(N_y\in 2\mathbb{Z}+1)\\
    2(4N_x+5)(N_y+1)-4&(N_y\in 2\mathbb{Z})
    \end{cases}.
\end{align}
When $N_y$ is odd, the $(010)$ zigzag edge meets the convex part of the armchair edges, while meeting the concave part when $N_y$ is even.
We consider the case $N_x=N_y$ in the following, and set $L=2N_x+1=2N_y+1$ to represent the number of hexagons in the horizontal and vertical directions.
{Thus, $N_{\rm site}=3566$ considered in the main text corresponds to $L=41$ and $N_x=N_y=20$.}

For general parameters, the Haldane model has the threefold rotation symmetry $C_{3z}$ and the magnetic mirror symmetry $\Theta M_x$ in the bulk.
When the next-nearest-neighbor hopping is pure imaginary, $\phi=\pi/2$, the model additionally has the rotation-particle-hole symmetry
\begin{subequations}\begin{gather}
    C_{2z}\mathcal{C}H^*(\mu)[C_{2z}\mathcal{C}]^{\dagger}=-H(-\mu),\\
    C_{2z}\mathcal{C}\ket{\bm{R}_s}=\ket{-\bm{R}_s}\times\begin{cases}
        +1&(s=A)\\
        -1&(s=B)
    \end{cases},
\end{gather}\end{subequations}
where the chemical potential dependence of the Hamiltonian is written explicitly.
Note that $\ket{-\bm{R}_s}$ belongs to the sublattice $-s$.
The flake geometry is invariant under the $C_{2z}$ operation $\bm{r}\to-\bm{r}$ as well as the $M_x$ operation $x\to-x$, and thus $\Theta M_x$ and $C_{2z}\mathcal{C}$ symmetries are exactly satisfied with OBCs when they are satisfied in the bulk.
While $C_{3z}$ symmetry is not preserved with OBCs, the local magnetization and current density approximately satisfy the $C_{3z}$ symmetry when the reference point is away from the boundaries, owing to the boundary insensitivity.

Note that $C_{2z}\mathcal{C}n(\bm{r})[C_{2z}\mathcal{C}]^{\dagger}=n(-\bm{r})$ and $\ket{\bm{r}}^*=\ket{\bm{r}}$ by definition, and thus
the bond-current operator satisfies
\begin{align}
    C_{2z}\mathcal{C}J_{\bm{r}\to \bm{r}'}[C_{2z}\mathcal{C}]^{\dagger}=-J^*_{-\bm{r}\to -\bm{r}'}.
\end{align}
This leads to the corresponding identity of the smeared current density operator
\begin{align}
    C_{2z}\mathcal{C}\bar{\bm{j}}(\bm{x})[C_{2z}\mathcal{C}]^{\dagger}=-\bar{\bm{j}}^*(-\bm{x}).
\end{align}
Thus, the smeared current density satisfies
\begin{align}
    \braket{\bar{\bm{j}}(\bm{x})}_\mu&\equiv \Tr[\bar{\bm{j}}(\bm{x})f(H(\mu))]\notag\\
    &=\Tr[C_{2z}\mathcal{C}\bar{\bm{j}}(\bm{x})[C_{2z}\mathcal{C}]^{\dagger}C_{2z}\mathcal{C}f(H(\mu))[C_{2z}\mathcal{C}]^{\dagger}]\notag\\
    &=-\Tr[\bar{\bm{j}}(-\bm{x})f(-H(-\mu))]^*\notag\\
    &=-\Tr[\bar{\bm{j}}(-\bm{x})]^*+\Tr[\bar{\bm{j}}(-\bm{x})f(H(-\mu))]^*\notag\\
    &=\braket{\bar{\bm{j}}(-\bm{x})}_{-\mu}.
\end{align}
This requires the net magnetic moment to flip when $\mu$ is inverted, and thus we obtain the vanishing bulk orbital magnetization $m_{\rm bulk}=0$ in the $\mu=0$ case.
When $\mu=0$, this requires local orbital magnetization to flip when $\bm{x}\to-\bm{x}$, to ensure the entire vanishing orbital magnetic moment.

\section{Numerical calculation details}
Here we show details of the numerical calculations.
We acknowledge the use of ChatGPT (OpenAI; GPT-5.4 Pro and GPT-5.5 Pro, accessed between March-June 2026) to assist with code generation. The generated codes were reviewed line by line by the author, and its outputs were cross-checked for consistency across different calculations and with analytical results. The author takes full responsibility for the code, the results, and the contents of the manuscript.

In Fig.1(b), NLM was evaluated based on its definition Eq.~\eqref{eq:m1_SM} for the A and B sites belonging to the home hexagon.
This is compared with the Bloch-state formula for the two-band model~\eqref{eq:LM_bulk_formula}.
Figures~1(c) and (d) are evaluated based on the definition~\eqref{eq:chi_defs_SM} and~\eqref{eq:magkernel_SM} of the three-point formula for inside the home hexagon.

In the following, we show the details of the other figures and additional calculation results.
We first introduce the local markers used for comparison~\cite{Bianco2013-vb,Seleznev2023-tz,Saati2025-hg}.
The BR-~\cite{Bianco2013-vb} and SV-~\cite{Seleznev2023-tz} local markers used in the $T=0$ numerical calculations are defined by
\begin{align}
     \mathcal{M}^{\rm BR}(\bm{r})&= 
\operatorname{Im}\Tr[n(\bm{r})
(-XHY^\dagger+X^\dagger HY)],\\
\mathcal{M}^{\rm SV}(\bm{r})&=\frac{1}{3}\Im\Tr[n(\bm{r})(-XHY^\dagger-HY^\dagger X-Y^\dagger XH\notag\\
&\qquad+X^\dagger HY+HYX^\dagger+YX^\dagger H)],
\end{align}
where $X\equiv P\hat{x}Q$, $Y\equiv P\hat{y}Q$, $Q=1-P$, and $P$ is the projection to the occupied states.
Note that $H$ includes the chemical potential $\mu$ in our notation.
The finite-temperature marker proposed by Saati-Hur-Pi\'echon (SHP) in Ref.~\cite{Saati2025-hg} is 
given by
\begin{subequations}\label{eq:Morb_local_SHP}
\begin{align}
\mathcal{M}_i^{\rm SHP}(\bm{r})
=
-
\sum_n
\Bigl[
f(\varepsilon_n)\, m_n^i(\bm{r})
-
F(\varepsilon_n)\, \Omega_n^i(\bm{r})
\Bigr],
\end{align}
with 
\begin{align}
\bm{m}_n(\bm{r})
&=
\left|\langle \bm{r} | n \rangle\right|^2
\, \bm{m}_n,\quad \bm{m}_n=-\braket{n|\frac{1}{2}\hat{\bm{r}}\times\bm{J}|n},
\\
\bm{\Omega}_n(\bm{r})
&=
\operatorname{Re}
\sum_{m\neq n}
\langle \bm{r} | n \rangle
\langle m | \bm{r} \rangle
\,
\frac{
\langle n |
\hat{\bm{r}}\times\hat{\bm{J}}
| m \rangle
}{
\varepsilon_n-\varepsilon_m
}.
\end{align}
\end{subequations}

We introduce the definitions of the NLM-related quantities evaluated in numerical calculations.
Analogous quantities are introduced for the other local markers by replacing the NLM with the local markers in the above.
The MQM from NLM is defined by
\begin{align}
    M_{\rm NLM}^{ij}&=\sum_{\bm{r}}r_i\mathcal{M}_j(\bm{r})-\frac{\delta_{ij}}{3}r_k{\mathcal{M}}_k(\bm{r}),
\end{align}
This is compared with the MQM defined from the bond currents [$M_{\rm bond}^{ij}$], evaluated by the first line of Eq.~\eqref{eq:Mij_SM_ave}.
Edge magnetization near the edge center is given by
\begin{align}
    \mathcal{M}_{(010)}\equiv\frac{1}{\sqrt{3}a}\sum_{\bm{r};x\sim0}\int_{0}^\infty dy\,\frac{e^{-\frac{(y-r_y)^2}{2l_W^2}}}{\sqrt{2\pi l_W^2}}{\mathcal{M}}_z(\bm{r}),
\end{align}
as discussed in End Matter.
In numerical calculations, we evaluated $[\mathcal{M}_{(010)}-\mathcal{M}_{(0\bar{1}0)}]/2$ with 
\begin{align}
    \mathcal{M}_{(0\bar{1}0)}\equiv\frac{1}{\sqrt{3}a}\sum_{\bm{r};x\sim0}\int^{0}_{-\infty} dy\,\frac{e^{-\frac{(y-r_y)^2}{2l_W^2}}}{\sqrt{2\pi l_W^2}}{\mathcal{M}}_z(\bm{r}).
\end{align}
This does not change the result since $\mathcal{M}_{(010)}=-\mathcal{M}_{(0\bar{1}0)}$ owing to the $C_{2z}\mathcal{C}$ symmetry.
The averaged edge magnetic moment is given by $M_{(010)}\equiv\frac{1}{D_x}\int_{-\infty}^\infty dx\int_{0}^\infty dy\,\bar{\mathcal{M}}_z(\bm{r})$ as discussed in the main text.
Here, $D_x$ corresponds to a system diameter in the $x$ direction and is defined by $D_x=D_y/V$, where $V=V_{\rm UC}N_{\rm UC}$ is the system area and $D_y\equiv\max_{\bm{r}}[r_y]-\min_{\bm{r}}[r_y]$.
This is also calculated by $[M_{(010)}-M_{(0\bar{1}0)}]/2$ with $M_{(0\bar{1}0)}\equiv\frac{1}{D_x}\int_{-\infty}^\infty dx\int^{0}_{-\infty} dy\,\bar{\mathcal{M}}_z(\bm{r})$.

The magnetization current from the local markers is evaluated by
\begin{align}
    \bm{j}_{\rm NLM}(\bm{x})\equiv \sum_{\bm{r}}[\nabla_{\bm{x}}W(\bm{x}-\bm{r})]\times \mathcal{\bm{M}}(\bm{r}),
\end{align}
rather than by directly performing a numerical derivative $\nabla_{\bm{x}}\times\bar{\mathcal{\bm{M}}}(\bm{x})$.
For the evaluation of the magnetization current density from the three-point formula, we used \begin{align}
    \bm{j}_{\rm 3pt}(\bm{x})&=\bm{\nabla}_{\bm{x}}\times\int d^d\bar{\bm{x}}\,W(\bm{x}-\bar{\bm{x}})m_{\rm orb}(\bar{\bm{x}})\notag\\
    &=\sum_{\bm{r}_1\bm{r}_2\bm{r}_3}m_{\bm{r}_1\bm{r}_2\bm{r}_3}\oint_{\partial\triangle_{\bm{r}_1\bm{r}_2\bm{r}_3}}d\bar{\bm{x}}\,W(\bm{x}-\bar{\bm{x}}),
\end{align}
by using the Stokes' theorem, to avoid numerically taking the $\bm{x}$ derivative.
The currents are compared with the expectation value of the smeared current density operator $\braket{\bar{\bm{j}}(\bm{x})}$, which is calculated by using the bond currents.

All the finite-size data related to MQM are fitted by using the data for $x\le0.01$ with the form such as
\begin{align}
   M^{yz}_{\rm orb}/V=a+b x+cx^2, \quad x=1/\sqrt{N_{\rm site}},
\end{align}
except for the BR-marker results which additionally required a cubic term $x^3$ and for the edge magnetization near the center $\mathcal{M}_{(010)}$ fitted by a constant.
Here, $V=V_{\rm UC}N_{\rm UC}$ is the system area.
The results are shown as solid or dashed curves in each figure.

\begin{figure}[t]
    \centering
    \includegraphics[width=\linewidth]{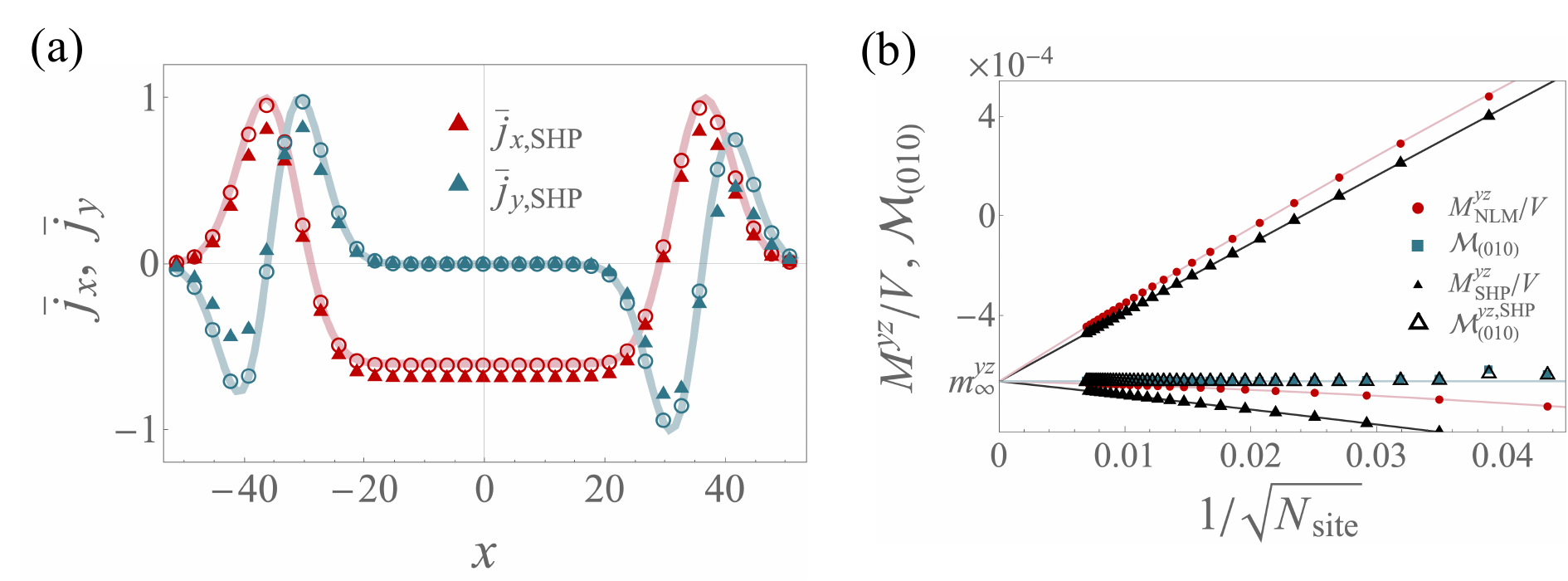}
    \caption{(a) Smeared current density and (b) size scaling of MQM-related quantities in the $T=0.1$ OQM state evaluated with the NLM and SHP formulas. The NLM results are the same as Fig.2(b) and (d).
    SHP results are shown by (a) closed red (blue) filled triangles for $\bar{j}_x$ ($\bar{j}_y$), and (b) filled (open) black triangles for $M^{yz}_{\rm SHP}/V$ ($\mathcal{M}_{(010)}^{yz,{\rm SHP}}$).
    }
    \label{figSHP}
\end{figure}

In Fig.2(a) and all the other figures, the Gaussian $W(\bm{x})\equiv\frac{1}{2\pi l_W^2}e^{-|\bm{x}|^2/2l_W^2}$ with $l_W=5a$ was used as the smearing function.
In Fig.2(b) (and Fig.~\ref{figSHP}(a) below), the currents and NLM are normalized by $\mathrm{max}_x|\bar{j}_x|=9.7\times 10^{-6}$, $\mathrm{max}_x|\bar{j}_y|=1.3\times 10^{-5}$, and $\mathrm{max}_x|\bar{\mathcal{M}}_z|=8.1\times 10^{-5}$, thus currents and NLM are shown in the range $[-1,1]$.
Here and hereafter, the $x$ scan was performed on the same line $y=26a$.
The maximum deviation of the smeared current densities by NLM from the bond-current results was less than $5\times 10^{-7}$, while the magnetization current evaluated from the three-point formula
coincides with the exact ones up to a numerical error of $\mathcal{O}(10^{-16})$, as they are analytically equivalent.

We show in Fig.~\ref{figSHP} the results of the SHP marker for the OQM state.
Figure~\ref{figSHP}(a) shows the current scan of the OQM state with the same settings as Fig.2(b), where the NLM and currents from three-point formula are abbreviated and instead the currents from SHP are additionally shown.
The smeared current density given as the magnetization current of the SHP marker [filled triangles] does not qualitatively reproduce the exact current density from the bond current [solid lines].
Figure~\ref{figSHP}(b) shows the size scaling of MQM and surface magnetization.
The SHP marker seems to reproduce MQM per unit area in the thermodynamic limit, but does not reproduce MQM in finite-size systems.
The edge magnetization defined from the SHP marker [open triangles] showed similar behavior to NLM.
In Fig.2(d) and Fig.~\ref{figSHP}(b), the upper and lower branches correspond to the concave [$N_y\in2\mathbb{Z}$] and convex-type [$N_y\in2\mathbb{Z}+1$] corner terminations, respectively.


\begin{thebibliography}{39}%
\makeatletter
\providecommand \@ifxundefined [1]{%
 \@ifx{#1\undefined}
}%
\providecommand \@ifnum [1]{%
 \ifnum #1\expandafter \@firstoftwo
 \else \expandafter \@secondoftwo
 \fi
}%
\providecommand \@ifx [1]{%
 \ifx #1\expandafter \@firstoftwo
 \else \expandafter \@secondoftwo
 \fi
}%
\providecommand \natexlab [1]{#1}%
\providecommand \enquote  [1]{``#1''}%
\providecommand \bibnamefont  [1]{#1}%
\providecommand \bibfnamefont [1]{#1}%
\providecommand \citenamefont [1]{#1}%
\providecommand \href@noop [0]{\@secondoftwo}%
\providecommand \href [0]{\begingroup \@sanitize@url \@href}%
\providecommand \@href[1]{\@@startlink{#1}\@@href}%
\providecommand \@@href[1]{\endgroup#1\@@endlink}%
\providecommand \@sanitize@url [0]{\catcode `\\12\catcode `\$12\catcode `\&12\catcode `\#12\catcode `\^12\catcode `\_12\catcode `\%12\relax}%
\providecommand \@@startlink[1]{}%
\providecommand \@@endlink[0]{}%
\providecommand \url  [0]{\begingroup\@sanitize@url \@url }%
\providecommand \@url [1]{\endgroup\@href {#1}{\urlprefix }}%
\providecommand \urlprefix  [0]{URL }%
\providecommand \Eprint [0]{\href }%
\providecommand \doibase [0]{https://doi.org/}%
\providecommand \selectlanguage [0]{\@gobble}%
\providecommand \bibinfo  [0]{\@secondoftwo}%
\providecommand \bibfield  [0]{\@secondoftwo}%
\providecommand \translation [1]{[#1]}%
\providecommand \BibitemOpen [0]{}%
\providecommand \bibitemStop [0]{}%
\providecommand \bibitemNoStop [0]{.\EOS\space}%
\providecommand \EOS [0]{\spacefactor3000\relax}%
\providecommand \BibitemShut  [1]{\csname bibitem#1\endcsname}%
\let\auto@bib@innerbib\@empty
\bibitem [{\citenamefont {Varma}(1997)}]{Varma1997-ho}%
  \BibitemOpen
  \bibfield  {author} {\bibinfo {author} {\bibfnamefont {C.~M.}\ \bibnamefont {Varma}},\ }\bibfield  {title} {\bibinfo {title} {Non-fermi-liquid states and pairing instability of a general model of copper oxide metals},\ }\href {https://doi.org/10.1103/PhysRevB.55.14554} {\bibfield  {journal} {\bibinfo  {journal} {Phys. Rev. B}\ }\textbf {\bibinfo {volume} {55}},\ \bibinfo {pages} {14554} (\bibinfo {year} {1997})}\BibitemShut {NoStop}%
\bibitem [{\citenamefont {Chakravarty}\ \emph {et~al.}(2001)\citenamefont {Chakravarty}, \citenamefont {Laughlin}, \citenamefont {Morr},\ and\ \citenamefont {Nayak}}]{Chakravarty2001-iu}%
  \BibitemOpen
  \bibfield  {author} {\bibinfo {author} {\bibfnamefont {S.}~\bibnamefont {Chakravarty}}, \bibinfo {author} {\bibfnamefont {R.~B.}\ \bibnamefont {Laughlin}}, \bibinfo {author} {\bibfnamefont {D.~K.}\ \bibnamefont {Morr}},\ and\ \bibinfo {author} {\bibfnamefont {C.}~\bibnamefont {Nayak}},\ }\bibfield  {title} {\bibinfo {title} {Hidden order in the cuprates},\ }\href {https://doi.org/10.1103/physrevb.63.094503} {\bibfield  {journal} {\bibinfo  {journal} {Phys. Rev. B}\ }\textbf {\bibinfo {volume} {63}},\ \bibinfo {pages} {094503} (\bibinfo {year} {2001})}\BibitemShut {NoStop}%
\bibitem [{\citenamefont {Resta}(2010)}]{Resta2010-if}%
  \BibitemOpen
  \bibfield  {author} {\bibinfo {author} {\bibfnamefont {R.}~\bibnamefont {Resta}},\ }\bibfield  {title} {\bibinfo {title} {Electrical polarization and orbital magnetization: the modern theories},\ }\href {https://doi.org/10.1088/0953-8984/22/12/123201} {\bibfield  {journal} {\bibinfo  {journal} {J. Phys. Condens. Matter}\ }\textbf {\bibinfo {volume} {22}},\ \bibinfo {pages} {123201} (\bibinfo {year} {2010})}\BibitemShut {NoStop}%
\bibitem [{\citenamefont {Tschirhart}\ \emph {et~al.}(2021)\citenamefont {Tschirhart}, \citenamefont {Serlin}, \citenamefont {Polshyn}, \citenamefont {Shragai}, \citenamefont {Xia}, \citenamefont {Zhu}, \citenamefont {Zhang}, \citenamefont {Watanabe}, \citenamefont {Taniguchi}, \citenamefont {Huber},\ and\ \citenamefont {Young}}]{Tschirhart2021-zh}%
  \BibitemOpen
  \bibfield  {author} {\bibinfo {author} {\bibfnamefont {C.~L.}\ \bibnamefont {Tschirhart}}, \bibinfo {author} {\bibfnamefont {M.}~\bibnamefont {Serlin}}, \bibinfo {author} {\bibfnamefont {H.}~\bibnamefont {Polshyn}}, \bibinfo {author} {\bibfnamefont {A.}~\bibnamefont {Shragai}}, \bibinfo {author} {\bibfnamefont {Z.}~\bibnamefont {Xia}}, \bibinfo {author} {\bibfnamefont {J.}~\bibnamefont {Zhu}}, \bibinfo {author} {\bibfnamefont {Y.}~\bibnamefont {Zhang}}, \bibinfo {author} {\bibfnamefont {K.}~\bibnamefont {Watanabe}}, \bibinfo {author} {\bibfnamefont {T.}~\bibnamefont {Taniguchi}}, \bibinfo {author} {\bibfnamefont {M.~E.}\ \bibnamefont {Huber}},\ and\ \bibinfo {author} {\bibfnamefont {A.~F.}\ \bibnamefont {Young}},\ }\bibfield  {title} {\bibinfo {title} {Imaging orbital ferromagnetism in a moir\'e chern insulator},\ }\href {https://doi.org/10.1126/science.abd3190} {\bibfield  {journal} {\bibinfo  {journal} {Science}\ }\textbf {\bibinfo {volume} {372}},\ \bibinfo {pages} {1323} (\bibinfo {year}
  {2021})}\BibitemShut {NoStop}%
\bibitem [{\citenamefont {Grover}\ \emph {et~al.}(2022)\citenamefont {Grover}, \citenamefont {Bocarsly}, \citenamefont {Uri}, \citenamefont {Stepanov}, \citenamefont {Di~Battista}, \citenamefont {Roy}, \citenamefont {Xiao}, \citenamefont {Meltzer}, \citenamefont {Myasoedov}, \citenamefont {Pareek}, \citenamefont {Watanabe}, \citenamefont {Taniguchi}, \citenamefont {Yan}, \citenamefont {Stern}, \citenamefont {Berg}, \citenamefont {Efetov},\ and\ \citenamefont {Zeldov}}]{Grover2022-mz}%
  \BibitemOpen
  \bibfield  {author} {\bibinfo {author} {\bibfnamefont {S.}~\bibnamefont {Grover}}, \bibinfo {author} {\bibfnamefont {M.}~\bibnamefont {Bocarsly}}, \bibinfo {author} {\bibfnamefont {A.}~\bibnamefont {Uri}}, \bibinfo {author} {\bibfnamefont {P.}~\bibnamefont {Stepanov}}, \bibinfo {author} {\bibfnamefont {G.}~\bibnamefont {Di~Battista}}, \bibinfo {author} {\bibfnamefont {I.}~\bibnamefont {Roy}}, \bibinfo {author} {\bibfnamefont {J.}~\bibnamefont {Xiao}}, \bibinfo {author} {\bibfnamefont {A.~Y.}\ \bibnamefont {Meltzer}}, \bibinfo {author} {\bibfnamefont {Y.}~\bibnamefont {Myasoedov}}, \bibinfo {author} {\bibfnamefont {K.}~\bibnamefont {Pareek}}, \bibinfo {author} {\bibfnamefont {K.}~\bibnamefont {Watanabe}}, \bibinfo {author} {\bibfnamefont {T.}~\bibnamefont {Taniguchi}}, \bibinfo {author} {\bibfnamefont {B.}~\bibnamefont {Yan}}, \bibinfo {author} {\bibfnamefont {A.}~\bibnamefont {Stern}}, \bibinfo {author} {\bibfnamefont {E.}~\bibnamefont {Berg}}, \bibinfo {author} {\bibfnamefont {D.~K.}\ \bibnamefont
  {Efetov}},\ and\ \bibinfo {author} {\bibfnamefont {E.}~\bibnamefont {Zeldov}},\ }\bibfield  {title} {\bibinfo {title} {Chern mosaic and berry-curvature magnetism in magic-angle graphene},\ }\href {https://doi.org/10.1038/s41567-022-01635-7} {\bibfield  {journal} {\bibinfo  {journal} {Nat. Phys.}\ }\textbf {\bibinfo {volume} {18}},\ \bibinfo {pages} {885} (\bibinfo {year} {2022})}\BibitemShut {NoStop}%
\bibitem [{\citenamefont {Han}\ \emph {et~al.}(2023)\citenamefont {Han}, \citenamefont {Lu}, \citenamefont {Scuri}, \citenamefont {Sung}, \citenamefont {Wang}, \citenamefont {Han}, \citenamefont {Watanabe}, \citenamefont {Taniguchi}, \citenamefont {Fu}, \citenamefont {Park},\ and\ \citenamefont {Ju}}]{Han2023-en}%
  \BibitemOpen
  \bibfield  {author} {\bibinfo {author} {\bibfnamefont {T.}~\bibnamefont {Han}}, \bibinfo {author} {\bibfnamefont {Z.}~\bibnamefont {Lu}}, \bibinfo {author} {\bibfnamefont {G.}~\bibnamefont {Scuri}}, \bibinfo {author} {\bibfnamefont {J.}~\bibnamefont {Sung}}, \bibinfo {author} {\bibfnamefont {J.}~\bibnamefont {Wang}}, \bibinfo {author} {\bibfnamefont {T.}~\bibnamefont {Han}}, \bibinfo {author} {\bibfnamefont {K.}~\bibnamefont {Watanabe}}, \bibinfo {author} {\bibfnamefont {T.}~\bibnamefont {Taniguchi}}, \bibinfo {author} {\bibfnamefont {L.}~\bibnamefont {Fu}}, \bibinfo {author} {\bibfnamefont {H.}~\bibnamefont {Park}},\ and\ \bibinfo {author} {\bibfnamefont {L.}~\bibnamefont {Ju}},\ }\bibfield  {title} {\bibinfo {title} {Orbital multiferroicity in pentalayer rhombohedral graphene},\ }\href {https://doi.org/10.1038/s41586-023-06572-w} {\bibfield  {journal} {\bibinfo  {journal} {Nature}\ }\textbf {\bibinfo {volume} {623}},\ \bibinfo {pages} {41} (\bibinfo {year} {2023})}\BibitemShut {NoStop}%
\bibitem [{\citenamefont {Jiang}\ \emph {et~al.}(2021)\citenamefont {Jiang}, \citenamefont {Yin}, \citenamefont {Denner}, \citenamefont {Shumiya}, \citenamefont {Ortiz}, \citenamefont {Xu}, \citenamefont {Guguchia}, \citenamefont {He}, \citenamefont {Hossain}, \citenamefont {Liu}, \citenamefont {Ruff}, \citenamefont {Kautzsch}, \citenamefont {Zhang}, \citenamefont {Chang}, \citenamefont {Belopolski}, \citenamefont {Zhang}, \citenamefont {Cochran}, \citenamefont {Multer}, \citenamefont {Litskevich}, \citenamefont {Cheng}, \citenamefont {Yang}, \citenamefont {Wang}, \citenamefont {Thomale}, \citenamefont {Neupert}, \citenamefont {Wilson},\ and\ \citenamefont {Hasan}}]{Jiang2021-lj}%
  \BibitemOpen
  \bibfield  {author} {\bibinfo {author} {\bibfnamefont {Y.-X.}\ \bibnamefont {Jiang}}, \bibinfo {author} {\bibfnamefont {J.-X.}\ \bibnamefont {Yin}}, \bibinfo {author} {\bibfnamefont {M.~M.}\ \bibnamefont {Denner}}, \bibinfo {author} {\bibfnamefont {N.}~\bibnamefont {Shumiya}}, \bibinfo {author} {\bibfnamefont {B.~R.}\ \bibnamefont {Ortiz}}, \bibinfo {author} {\bibfnamefont {G.}~\bibnamefont {Xu}}, \bibinfo {author} {\bibfnamefont {Z.}~\bibnamefont {Guguchia}}, \bibinfo {author} {\bibfnamefont {J.}~\bibnamefont {He}}, \bibinfo {author} {\bibfnamefont {M.~S.}\ \bibnamefont {Hossain}}, \bibinfo {author} {\bibfnamefont {X.}~\bibnamefont {Liu}}, \bibinfo {author} {\bibfnamefont {J.}~\bibnamefont {Ruff}}, \bibinfo {author} {\bibfnamefont {L.}~\bibnamefont {Kautzsch}}, \bibinfo {author} {\bibfnamefont {S.~S.}\ \bibnamefont {Zhang}}, \bibinfo {author} {\bibfnamefont {G.}~\bibnamefont {Chang}}, \bibinfo {author} {\bibfnamefont {I.}~\bibnamefont {Belopolski}}, \bibinfo {author} {\bibfnamefont {Q.}~\bibnamefont
  {Zhang}}, \bibinfo {author} {\bibfnamefont {T.~A.}\ \bibnamefont {Cochran}}, \bibinfo {author} {\bibfnamefont {D.}~\bibnamefont {Multer}}, \bibinfo {author} {\bibfnamefont {M.}~\bibnamefont {Litskevich}}, \bibinfo {author} {\bibfnamefont {Z.-J.}\ \bibnamefont {Cheng}}, \bibinfo {author} {\bibfnamefont {X.~P.}\ \bibnamefont {Yang}}, \bibinfo {author} {\bibfnamefont {Z.}~\bibnamefont {Wang}}, \bibinfo {author} {\bibfnamefont {R.}~\bibnamefont {Thomale}}, \bibinfo {author} {\bibfnamefont {T.}~\bibnamefont {Neupert}}, \bibinfo {author} {\bibfnamefont {S.~D.}\ \bibnamefont {Wilson}},\ and\ \bibinfo {author} {\bibfnamefont {M.~Z.}\ \bibnamefont {Hasan}},\ }\bibfield  {title} {\bibinfo {title} {Unconventional chiral charge order in kagome superconductor {KV3Sb5}},\ }\href {https://doi.org/10.1038/s41563-021-01034-y} {\bibfield  {journal} {\bibinfo  {journal} {Nat. Mater.}\ }\textbf {\bibinfo {volume} {20}},\ \bibinfo {pages} {1353} (\bibinfo {year} {2021})}\BibitemShut {NoStop}%
\bibitem [{\citenamefont {Mielke}\ \emph {et~al.}(2022)\citenamefont {Mielke}, \citenamefont {Das}, \citenamefont {Yin}, \citenamefont {Liu}, \citenamefont {Gupta}, \citenamefont {Jiang}, \citenamefont {Medarde}, \citenamefont {Wu}, \citenamefont {Lei}, \citenamefont {Chang}, \citenamefont {Dai}, \citenamefont {Si}, \citenamefont {Miao}, \citenamefont {Thomale}, \citenamefont {Neupert}, \citenamefont {Shi}, \citenamefont {Khasanov}, \citenamefont {Hasan}, \citenamefont {Luetkens},\ and\ \citenamefont {Guguchia}}]{Mielke2022-gs}%
  \BibitemOpen
  \bibfield  {author} {\bibinfo {author} {\bibfnamefont {C.}~\bibnamefont {Mielke}, \bibfnamefont {3rd}}, \bibinfo {author} {\bibfnamefont {D.}~\bibnamefont {Das}}, \bibinfo {author} {\bibfnamefont {J.-X.}\ \bibnamefont {Yin}}, \bibinfo {author} {\bibfnamefont {H.}~\bibnamefont {Liu}}, \bibinfo {author} {\bibfnamefont {R.}~\bibnamefont {Gupta}}, \bibinfo {author} {\bibfnamefont {Y.-X.}\ \bibnamefont {Jiang}}, \bibinfo {author} {\bibfnamefont {M.}~\bibnamefont {Medarde}}, \bibinfo {author} {\bibfnamefont {X.}~\bibnamefont {Wu}}, \bibinfo {author} {\bibfnamefont {H.~C.}\ \bibnamefont {Lei}}, \bibinfo {author} {\bibfnamefont {J.}~\bibnamefont {Chang}}, \bibinfo {author} {\bibfnamefont {P.}~\bibnamefont {Dai}}, \bibinfo {author} {\bibfnamefont {Q.}~\bibnamefont {Si}}, \bibinfo {author} {\bibfnamefont {H.}~\bibnamefont {Miao}}, \bibinfo {author} {\bibfnamefont {R.}~\bibnamefont {Thomale}}, \bibinfo {author} {\bibfnamefont {T.}~\bibnamefont {Neupert}}, \bibinfo {author} {\bibfnamefont {Y.}~\bibnamefont {Shi}},
  \bibinfo {author} {\bibfnamefont {R.}~\bibnamefont {Khasanov}}, \bibinfo {author} {\bibfnamefont {M.~Z.}\ \bibnamefont {Hasan}}, \bibinfo {author} {\bibfnamefont {H.}~\bibnamefont {Luetkens}},\ and\ \bibinfo {author} {\bibfnamefont {Z.}~\bibnamefont {Guguchia}},\ }\bibfield  {title} {\bibinfo {title} {Time-reversal symmetry-breaking charge order in a kagome superconductor},\ }\href {https://doi.org/10.1038/s41586-021-04327-z} {\bibfield  {journal} {\bibinfo  {journal} {Nature}\ }\textbf {\bibinfo {volume} {602}},\ \bibinfo {pages} {245} (\bibinfo {year} {2022})}\BibitemShut {NoStop}%
\bibitem [{\citenamefont {Hirst}(1997)}]{Hirst1997-ei}%
  \BibitemOpen
  \bibfield  {author} {\bibinfo {author} {\bibfnamefont {L.~L.}\ \bibnamefont {Hirst}},\ }\bibfield  {title} {\bibinfo {title} {The microscopic magnetization: concept and application},\ }\href {https://doi.org/10.1103/RevModPhys.69.607} {\bibfield  {journal} {\bibinfo  {journal} {Rev. Mod. Phys.}\ }\textbf {\bibinfo {volume} {69}},\ \bibinfo {pages} {607} (\bibinfo {year} {1997})}\BibitemShut {NoStop}%
\bibitem [{\citenamefont {Xiao}\ \emph {et~al.}(2010)\citenamefont {Xiao}, \citenamefont {Chang},\ and\ \citenamefont {Niu}}]{Xiao2010-ah}%
  \BibitemOpen
  \bibfield  {author} {\bibinfo {author} {\bibfnamefont {D.}~\bibnamefont {Xiao}}, \bibinfo {author} {\bibfnamefont {M.-C.}\ \bibnamefont {Chang}},\ and\ \bibinfo {author} {\bibfnamefont {Q.}~\bibnamefont {Niu}},\ }\bibfield  {title} {\bibinfo {title} {Berry phase effects on electronic properties},\ }\href {https://doi.org/10.1103/RevModPhys.82.1959} {\bibfield  {journal} {\bibinfo  {journal} {Rev. Mod. Phys.}\ }\textbf {\bibinfo {volume} {82}},\ \bibinfo {pages} {1959} (\bibinfo {year} {2010})}\BibitemShut {NoStop}%
\bibitem [{\citenamefont {Xiao}\ \emph {et~al.}(2005)\citenamefont {Xiao}, \citenamefont {Shi},\ and\ \citenamefont {Niu}}]{Xiao2005-px}%
  \BibitemOpen
  \bibfield  {author} {\bibinfo {author} {\bibfnamefont {D.}~\bibnamefont {Xiao}}, \bibinfo {author} {\bibfnamefont {J.}~\bibnamefont {Shi}},\ and\ \bibinfo {author} {\bibfnamefont {Q.}~\bibnamefont {Niu}},\ }\bibfield  {title} {\bibinfo {title} {Berry phase correction to electron density of states in solids},\ }\href {https://doi.org/10.1103/PhysRevLett.95.137204} {\bibfield  {journal} {\bibinfo  {journal} {Phys. Rev. Lett.}\ }\textbf {\bibinfo {volume} {95}},\ \bibinfo {pages} {137204} (\bibinfo {year} {2005})}\BibitemShut {NoStop}%
\bibitem [{\citenamefont {Thonhauser}\ \emph {et~al.}(2005)\citenamefont {Thonhauser}, \citenamefont {Ceresoli}, \citenamefont {Vanderbilt},\ and\ \citenamefont {Resta}}]{Thonhauser2005-gi}%
  \BibitemOpen
  \bibfield  {author} {\bibinfo {author} {\bibfnamefont {T.}~\bibnamefont {Thonhauser}}, \bibinfo {author} {\bibfnamefont {D.}~\bibnamefont {Ceresoli}}, \bibinfo {author} {\bibfnamefont {D.}~\bibnamefont {Vanderbilt}},\ and\ \bibinfo {author} {\bibfnamefont {R.}~\bibnamefont {Resta}},\ }\bibfield  {title} {\bibinfo {title} {Orbital magnetization in periodic insulators},\ }\href {https://doi.org/10.1103/PhysRevLett.95.137205} {\bibfield  {journal} {\bibinfo  {journal} {Phys. Rev. Lett.}\ }\textbf {\bibinfo {volume} {95}},\ \bibinfo {pages} {137205} (\bibinfo {year} {2005})}\BibitemShut {NoStop}%
\bibitem [{\citenamefont {Ceresoli}\ \emph {et~al.}(2006)\citenamefont {Ceresoli}, \citenamefont {Thonhauser}, \citenamefont {Vanderbilt},\ and\ \citenamefont {Resta}}]{Ceresoli2006-fs}%
  \BibitemOpen
  \bibfield  {author} {\bibinfo {author} {\bibfnamefont {D.}~\bibnamefont {Ceresoli}}, \bibinfo {author} {\bibfnamefont {T.}~\bibnamefont {Thonhauser}}, \bibinfo {author} {\bibfnamefont {D.}~\bibnamefont {Vanderbilt}},\ and\ \bibinfo {author} {\bibfnamefont {R.}~\bibnamefont {Resta}},\ }\bibfield  {title} {\bibinfo {title} {Orbital magnetization in crystalline solids: Multi-band insulators, chern insulators, and metals},\ }\href {https://doi.org/10.1103/physrevb.74.024408} {\bibfield  {journal} {\bibinfo  {journal} {Phys. Rev. B}\ }\textbf {\bibinfo {volume} {74}},\ \bibinfo {pages} {024408} (\bibinfo {year} {2006})}\BibitemShut {NoStop}%
\bibitem [{\citenamefont {Shi}\ \emph {et~al.}(2007)\citenamefont {Shi}, \citenamefont {Vignale}, \citenamefont {Xiao},\ and\ \citenamefont {Niu}}]{Shi2007-rc}%
  \BibitemOpen
  \bibfield  {author} {\bibinfo {author} {\bibfnamefont {J.}~\bibnamefont {Shi}}, \bibinfo {author} {\bibfnamefont {G.}~\bibnamefont {Vignale}}, \bibinfo {author} {\bibfnamefont {D.}~\bibnamefont {Xiao}},\ and\ \bibinfo {author} {\bibfnamefont {Q.}~\bibnamefont {Niu}},\ }\bibfield  {title} {\bibinfo {title} {Quantum theory of orbital magnetization and its generalization to interacting systems},\ }\href {https://doi.org/10.1103/PhysRevLett.99.197202} {\bibfield  {journal} {\bibinfo  {journal} {Phys. Rev. Lett.}\ }\textbf {\bibinfo {volume} {99}},\ \bibinfo {pages} {197202} (\bibinfo {year} {2007})}\BibitemShut {NoStop}%
\bibitem [{\citenamefont {Lopez}\ \emph {et~al.}(2012)\citenamefont {Lopez}, \citenamefont {Vanderbilt}, \citenamefont {Thonhauser},\ and\ \citenamefont {Souza}}]{Lopez2012-ut}%
  \BibitemOpen
  \bibfield  {author} {\bibinfo {author} {\bibfnamefont {M.~G.}\ \bibnamefont {Lopez}}, \bibinfo {author} {\bibfnamefont {D.}~\bibnamefont {Vanderbilt}}, \bibinfo {author} {\bibfnamefont {T.}~\bibnamefont {Thonhauser}},\ and\ \bibinfo {author} {\bibfnamefont {I.}~\bibnamefont {Souza}},\ }\bibfield  {title} {\bibinfo {title} {Wannier-based calculation of the orbital magnetization in crystals},\ }\href {https://doi.org/10.1103/physrevb.85.014435} {\bibfield  {journal} {\bibinfo  {journal} {Phys. Rev. B}\ }\textbf {\bibinfo {volume} {85}},\ \bibinfo {pages} {014435} (\bibinfo {year} {2012})}\BibitemShut {NoStop}%
\bibitem [{\citenamefont {Zhu}\ \emph {et~al.}(2012)\citenamefont {Zhu}, \citenamefont {Yang}, \citenamefont {Fang}, \citenamefont {Liu},\ and\ \citenamefont {Yao}}]{Zhu2012-hc}%
  \BibitemOpen
  \bibfield  {author} {\bibinfo {author} {\bibfnamefont {G.}~\bibnamefont {Zhu}}, \bibinfo {author} {\bibfnamefont {S.~A.}\ \bibnamefont {Yang}}, \bibinfo {author} {\bibfnamefont {C.}~\bibnamefont {Fang}}, \bibinfo {author} {\bibfnamefont {W.~M.}\ \bibnamefont {Liu}},\ and\ \bibinfo {author} {\bibfnamefont {Y.}~\bibnamefont {Yao}},\ }\bibfield  {title} {\bibinfo {title} {Theory of orbital magnetization in disordered systems},\ }\href {https://doi.org/10.1103/physrevb.86.214415} {\bibfield  {journal} {\bibinfo  {journal} {Phys. Rev. B}\ }\textbf {\bibinfo {volume} {86}},\ \bibinfo {pages} {214415} (\bibinfo {year} {2012})}\BibitemShut {NoStop}%
\bibitem [{\citenamefont {Nourafkan}\ \emph {et~al.}(2014)\citenamefont {Nourafkan}, \citenamefont {Kotliar},\ and\ \citenamefont {Tremblay}}]{Nourafkan2014-ju}%
  \BibitemOpen
  \bibfield  {author} {\bibinfo {author} {\bibfnamefont {R.}~\bibnamefont {Nourafkan}}, \bibinfo {author} {\bibfnamefont {G.}~\bibnamefont {Kotliar}},\ and\ \bibinfo {author} {\bibfnamefont {A.-M.~S.}\ \bibnamefont {Tremblay}},\ }\bibfield  {title} {\bibinfo {title} {Orbital magnetization of correlated electrons with arbitrary band topology},\ }\href {https://doi.org/10.1103/physrevb.90.125132} {\bibfield  {journal} {\bibinfo  {journal} {Phys. Rev. B}\ }\textbf {\bibinfo {volume} {90}},\ \bibinfo {pages} {125132} (\bibinfo {year} {2014})}\BibitemShut {NoStop}%
\bibitem [{\citenamefont {Bianco}\ and\ \citenamefont {Resta}(2013)}]{Bianco2013-vb}%
  \BibitemOpen
  \bibfield  {author} {\bibinfo {author} {\bibfnamefont {R.}~\bibnamefont {Bianco}}\ and\ \bibinfo {author} {\bibfnamefont {R.}~\bibnamefont {Resta}},\ }\bibfield  {title} {\bibinfo {title} {Orbital magnetization as a local property},\ }\href {https://doi.org/10.1103/PhysRevLett.110.087202} {\bibfield  {journal} {\bibinfo  {journal} {Phys. Rev. Lett.}\ }\textbf {\bibinfo {volume} {110}},\ \bibinfo {pages} {087202} (\bibinfo {year} {2013})}\BibitemShut {NoStop}%
\bibitem [{\citenamefont {Bianco}\ and\ \citenamefont {Resta}(2016)}]{Bianco2016-yk}%
  \BibitemOpen
  \bibfield  {author} {\bibinfo {author} {\bibfnamefont {R.}~\bibnamefont {Bianco}}\ and\ \bibinfo {author} {\bibfnamefont {R.}~\bibnamefont {Resta}},\ }\bibfield  {title} {\bibinfo {title} {Orbital magnetization in insulators: Bulk versus surface},\ }\href {https://doi.org/10.1103/physrevb.93.174417} {\bibfield  {journal} {\bibinfo  {journal} {Phys. Rev. B}\ }\textbf {\bibinfo {volume} {93}},\ \bibinfo {pages} {174417} (\bibinfo {year} {2016})}\BibitemShut {NoStop}%
\bibitem [{\citenamefont {Seleznev}\ and\ \citenamefont {Vanderbilt}(2023)}]{Seleznev2023-tz}%
  \BibitemOpen
  \bibfield  {author} {\bibinfo {author} {\bibfnamefont {D.}~\bibnamefont {Seleznev}}\ and\ \bibinfo {author} {\bibfnamefont {D.}~\bibnamefont {Vanderbilt}},\ }\bibfield  {title} {\bibinfo {title} {Towards a theory of surface orbital magnetization},\ }\href {https://doi.org/10.1103/physrevb.107.115102} {\bibfield  {journal} {\bibinfo  {journal} {Phys. Rev. B}\ }\textbf {\bibinfo {volume} {107}},\ \bibinfo {pages} {115102} (\bibinfo {year} {2023})}\BibitemShut {NoStop}%
\bibitem [{\citenamefont {Saati}\ \emph {et~al.}(2025)\citenamefont {Saati}, \citenamefont {Hur},\ and\ \citenamefont {Pi\'echon}}]{Saati2025-hg}%
  \BibitemOpen
  \bibfield  {author} {\bibinfo {author} {\bibfnamefont {S.~A.}\ \bibnamefont {Saati}}, \bibinfo {author} {\bibfnamefont {K.~L.}\ \bibnamefont {Hur}},\ and\ \bibinfo {author} {\bibfnamefont {F.}~\bibnamefont {Pi\'echon}},\ }\bibfield  {title} {\bibinfo {title} {Theory of local orbital magnetization: local berry curvature},\ }\bibfield  {journal} {\bibinfo  {journal} {arXiv [cond-mat.mes-hall]}\ }\href {https://doi.org/10.48550/arXiv.2512.12343} {10.48550/arXiv.2512.12343} (\bibinfo {year} {2025}),\ \Eprint {https://arxiv.org/abs/2512.12343} {arXiv:2512.12343 [cond-mat.mes-hall]} \BibitemShut {NoStop}%
\bibitem [{\citenamefont {Mahon}\ \emph {et~al.}(2023)\citenamefont {Mahon}, \citenamefont {Kattan},\ and\ \citenamefont {Sipe}}]{Mahon2023-dk}%
  \BibitemOpen
  \bibfield  {author} {\bibinfo {author} {\bibfnamefont {P.~T.}\ \bibnamefont {Mahon}}, \bibinfo {author} {\bibfnamefont {J.~G.}\ \bibnamefont {Kattan}},\ and\ \bibinfo {author} {\bibfnamefont {J.~E.}\ \bibnamefont {Sipe}},\ }\bibfield  {title} {\bibinfo {title} {Polarization and orbital magnetization in chern insulators: A microscopic perspective},\ }\href {https://doi.org/10.1103/physrevb.107.115110} {\bibfield  {journal} {\bibinfo  {journal} {Phys. Rev. B}\ }\textbf {\bibinfo {volume} {107}},\ \bibinfo {pages} {115110} (\bibinfo {year} {2023})}\BibitemShut {NoStop}%
\bibitem [{\citenamefont {Swiecicki}\ and\ \citenamefont {Sipe}(2014)}]{Swiecicki2014-nb}%
  \BibitemOpen
  \bibfield  {author} {\bibinfo {author} {\bibfnamefont {S.~D.}\ \bibnamefont {Swiecicki}}\ and\ \bibinfo {author} {\bibfnamefont {J.~E.}\ \bibnamefont {Sipe}},\ }\bibfield  {title} {\bibinfo {title} {Linear response of crystals to electromagnetic fields: Microscopic charge-current density, polarization, and magnetization},\ }\href {https://doi.org/10.1103/physrevb.90.125115} {\bibfield  {journal} {\bibinfo  {journal} {Phys. Rev. B}\ }\textbf {\bibinfo {volume} {90}},\ \bibinfo {pages} {125115} (\bibinfo {year} {2014})}\BibitemShut {NoStop}%
\bibitem [{\citenamefont {Mahon}\ \emph {et~al.}(2019)\citenamefont {Mahon}, \citenamefont {Muniz},\ and\ \citenamefont {Sipe}}]{Mahon2019-hd}%
  \BibitemOpen
  \bibfield  {author} {\bibinfo {author} {\bibfnamefont {P.~T.}\ \bibnamefont {Mahon}}, \bibinfo {author} {\bibfnamefont {R.~A.}\ \bibnamefont {Muniz}},\ and\ \bibinfo {author} {\bibfnamefont {J.~E.}\ \bibnamefont {Sipe}},\ }\bibfield  {title} {\bibinfo {title} {Microscopic polarization and magnetization fields in extended systems},\ }\href {https://doi.org/10.1103/physrevb.99.235140} {\bibfield  {journal} {\bibinfo  {journal} {Phys. Rev. B}\ }\textbf {\bibinfo {volume} {99}},\ \bibinfo {pages} {235140} (\bibinfo {year} {2019})}\BibitemShut {NoStop}%
\bibitem [{\citenamefont {Mahon}\ and\ \citenamefont {Sipe}(2022)}]{Mahon2022-ye}%
  \BibitemOpen
  \bibfield  {author} {\bibinfo {author} {\bibfnamefont {P.~T.}\ \bibnamefont {Mahon}}\ and\ \bibinfo {author} {\bibfnamefont {J.~E.}\ \bibnamefont {Sipe}},\ }\href@noop {} {\bibinfo {title} {Electric polarization and magnetization in metals}},\ \bibinfo {howpublished} {\url{https://scipost.org/preprints/scipost\_202206\_00031v3/}} (\bibinfo {year} {2022}),\ \bibinfo {note} {accessed: 2026-6-5}\BibitemShut {NoStop}%
\bibitem [{Sup()}]{Supplemental}%
  \BibitemOpen
  \href@noop {} {}\bibinfo {note} {See Supplemental Material for more details.}\BibitemShut {Stop}%
\bibitem [{\citenamefont {Haldane}(1988)}]{Haldane1988-og}%
  \BibitemOpen
  \bibfield  {author} {\bibinfo {author} {\bibfnamefont {F.~D.}\ \bibnamefont {Haldane}},\ }\bibfield  {title} {\bibinfo {title} {Model for a quantum hall effect without landau levels: Condensed-matter realization of the ``parity anomaly''},\ }\href {https://doi.org/10.1103/PhysRevLett.61.2015} {\bibfield  {journal} {\bibinfo  {journal} {Phys. Rev. Lett.}\ }\textbf {\bibinfo {volume} {61}},\ \bibinfo {pages} {2015} (\bibinfo {year} {1988})}\BibitemShut {NoStop}%
\bibitem [{\citenamefont {Chen}\ and\ \citenamefont {Lee}(2011)}]{Chen2011-dj}%
  \BibitemOpen
  \bibfield  {author} {\bibinfo {author} {\bibfnamefont {K.-T.}\ \bibnamefont {Chen}}\ and\ \bibinfo {author} {\bibfnamefont {P.~A.}\ \bibnamefont {Lee}},\ }\bibfield  {title} {\bibinfo {title} {Unified formalism for calculating polarization, magnetization, and more in a periodic insulator},\ }\href {https://doi.org/10.1103/physrevb.84.205137} {\bibfield  {journal} {\bibinfo  {journal} {Phys. Rev. B}\ }\textbf {\bibinfo {volume} {84}},\ \bibinfo {pages} {205137} (\bibinfo {year} {2011})}\BibitemShut {NoStop}%
\bibitem [{End()}]{EndMatter}%
  \BibitemOpen
  \href@noop {} {}\bibinfo {note} {See End Matter for more details.}\BibitemShut {Stop}%
\bibitem [{Note1()}]{Note1}%
  \BibitemOpen
  \bibinfo {note} {For $\epsilon _m=\epsilon _n$, $F_{mn}/\epsilon _{mn}\equiv \partial _{\epsilon _m}F(\epsilon _m)$}\BibitemShut {NoStop}%
\bibitem [{\citenamefont {Bianco}\ and\ \citenamefont {Resta}(2011)}]{Bianco2011-tv}%
  \BibitemOpen
  \bibfield  {author} {\bibinfo {author} {\bibfnamefont {R.}~\bibnamefont {Bianco}}\ and\ \bibinfo {author} {\bibfnamefont {R.}~\bibnamefont {Resta}},\ }\bibfield  {title} {\bibinfo {title} {Mapping topological order in coordinate space},\ }\href {https://doi.org/10.1103/physrevb.84.241106} {\bibfield  {journal} {\bibinfo  {journal} {Phys. Rev. B}\ }\textbf {\bibinfo {volume} {84}},\ \bibinfo {pages} {241106} (\bibinfo {year} {2011})}\BibitemShut {NoStop}%
\bibitem [{\citenamefont {Zhu}\ \emph {et~al.}(2021)\citenamefont {Zhu}, \citenamefont {Hughes},\ and\ \citenamefont {Alexandradinata}}]{Zhu2021-bo}%
  \BibitemOpen
  \bibfield  {author} {\bibinfo {author} {\bibfnamefont {P.}~\bibnamefont {Zhu}}, \bibinfo {author} {\bibfnamefont {T.~L.}\ \bibnamefont {Hughes}},\ and\ \bibinfo {author} {\bibfnamefont {A.}~\bibnamefont {Alexandradinata}},\ }\bibfield  {title} {\bibinfo {title} {Quantized surface magnetism and higher-order topology: Application to the hopf insulator},\ }\href {https://doi.org/10.1103/physrevb.103.014417} {\bibfield  {journal} {\bibinfo  {journal} {Phys. Rev. B}\ }\textbf {\bibinfo {volume} {103}},\ \bibinfo {pages} {014417} (\bibinfo {year} {2021})}\BibitemShut {NoStop}%
\bibitem [{\citenamefont {Gliozzi}\ \emph {et~al.}(2022)\citenamefont {Gliozzi}, \citenamefont {Lin},\ and\ \citenamefont {Hughes}}]{Gliozzi2022-gg}%
  \BibitemOpen
  \bibfield  {author} {\bibinfo {author} {\bibfnamefont {J.}~\bibnamefont {Gliozzi}}, \bibinfo {author} {\bibfnamefont {M.}~\bibnamefont {Lin}},\ and\ \bibinfo {author} {\bibfnamefont {T.~L.}\ \bibnamefont {Hughes}},\ }\bibfield  {title} {\bibinfo {title} {Orbital magnetic quadrupole moment in higher order topological phases},\ }\bibfield  {journal} {\bibinfo  {journal} {arXiv [cond-mat.mes-hall]}\ }\href {https://doi.org/10.48550/arXiv.2211.08438} {10.48550/arXiv.2211.08438} (\bibinfo {year} {2022}),\ \Eprint {https://arxiv.org/abs/2211.08438} {arXiv:2211.08438 [cond-mat.mes-hall]} \BibitemShut {NoStop}%
\bibitem [{\citenamefont {Jo}\ \emph {et~al.}(2024)\citenamefont {Jo}, \citenamefont {Go}, \citenamefont {Choi},\ and\ \citenamefont {Lee}}]{Jo2024-le}%
  \BibitemOpen
  \bibfield  {author} {\bibinfo {author} {\bibfnamefont {D.}~\bibnamefont {Jo}}, \bibinfo {author} {\bibfnamefont {D.}~\bibnamefont {Go}}, \bibinfo {author} {\bibfnamefont {G.-M.}\ \bibnamefont {Choi}},\ and\ \bibinfo {author} {\bibfnamefont {H.-W.}\ \bibnamefont {Lee}},\ }\bibfield  {title} {\bibinfo {title} {Spintronics meets orbitronics: Emergence of orbital angular momentum in solids},\ }\href {https://doi.org/10.1038/s44306-024-00023-6} {\bibfield  {journal} {\bibinfo  {journal} {Npj Spintron.}\ }\textbf {\bibinfo {volume} {2}},\ \bibinfo {pages} {19} (\bibinfo {year} {2024})}\BibitemShut {NoStop}%
\bibitem [{\citenamefont {Cysne}\ \emph {et~al.}(2025)\citenamefont {Cysne}, \citenamefont {Canonico}, \citenamefont {Costa}, \citenamefont {Muniz},\ and\ \citenamefont {Rappoport}}]{Cysne2025-ra}%
  \BibitemOpen
  \bibfield  {author} {\bibinfo {author} {\bibfnamefont {T.~P.}\ \bibnamefont {Cysne}}, \bibinfo {author} {\bibfnamefont {L.~M.}\ \bibnamefont {Canonico}}, \bibinfo {author} {\bibfnamefont {M.}~\bibnamefont {Costa}}, \bibinfo {author} {\bibfnamefont {R.~B.}\ \bibnamefont {Muniz}},\ and\ \bibinfo {author} {\bibfnamefont {T.~G.}\ \bibnamefont {Rappoport}},\ }\bibfield  {title} {\bibinfo {title} {Orbitronics in two-dimensional materials},\ }\href {https://doi.org/10.1038/s44306-025-00103-1} {\bibfield  {journal} {\bibinfo  {journal} {Npj Spintron.}\ }\textbf {\bibinfo {volume} {3}},\ \bibinfo {pages} {39} (\bibinfo {year} {2025})}\BibitemShut {NoStop}%
\bibitem [{\citenamefont {Essin}\ \emph {et~al.}(2010)\citenamefont {Essin}, \citenamefont {Turner}, \citenamefont {Moore},\ and\ \citenamefont {Vanderbilt}}]{Essin2010-jr}%
  \BibitemOpen
  \bibfield  {author} {\bibinfo {author} {\bibfnamefont {A.~M.}\ \bibnamefont {Essin}}, \bibinfo {author} {\bibfnamefont {A.~M.}\ \bibnamefont {Turner}}, \bibinfo {author} {\bibfnamefont {J.~E.}\ \bibnamefont {Moore}},\ and\ \bibinfo {author} {\bibfnamefont {D.}~\bibnamefont {Vanderbilt}},\ }\bibfield  {title} {\bibinfo {title} {Orbital magnetoelectric coupling in band insulators},\ }\href {https://doi.org/10.1103/physrevb.81.205104} {\bibfield  {journal} {\bibinfo  {journal} {Phys. Rev. B}\ }\textbf {\bibinfo {volume} {81}},\ \bibinfo {pages} {205104} (\bibinfo {year} {2010})}\BibitemShut {NoStop}%
\bibitem [{\citenamefont {Luttinger}(1951)}]{Luttinger1951-hj}%
  \BibitemOpen
  \bibfield  {author} {\bibinfo {author} {\bibfnamefont {J.~M.}\ \bibnamefont {Luttinger}},\ }\bibfield  {title} {\bibinfo {title} {The effect of a magnetic field on electrons in a periodic potential},\ }\href {https://doi.org/10.1103/PhysRev.84.814} {\bibfield  {journal} {\bibinfo  {journal} {Phys. Rev.}\ }\textbf {\bibinfo {volume} {84}},\ \bibinfo {pages} {814} (\bibinfo {year} {1951})}\BibitemShut {NoStop}%
\bibitem [{\citenamefont {Gao}\ and\ \citenamefont {Xiao}(2018)}]{Gao2018-cx}%
  \BibitemOpen
  \bibfield  {author} {\bibinfo {author} {\bibfnamefont {Y.}~\bibnamefont {Gao}}\ and\ \bibinfo {author} {\bibfnamefont {D.}~\bibnamefont {Xiao}},\ }\bibfield  {title} {\bibinfo {title} {Orbital magnetic quadrupole moment and nonlinear anomalous thermoelectric transport},\ }\href {https://doi.org/10.1103/physrevb.98.060402} {\bibfield  {journal} {\bibinfo  {journal} {Phys. Rev. B}\ }\textbf {\bibinfo {volume} {98}},\ \bibinfo {pages} {060402} (\bibinfo {year} {2018})}\BibitemShut {NoStop}%
\bibitem [{\citenamefont {Shitade}\ \emph {et~al.}(2018)\citenamefont {Shitade}, \citenamefont {Watanabe},\ and\ \citenamefont {Yanase}}]{Shitade2018-lo}%
  \BibitemOpen
  \bibfield  {author} {\bibinfo {author} {\bibfnamefont {A.}~\bibnamefont {Shitade}}, \bibinfo {author} {\bibfnamefont {H.}~\bibnamefont {Watanabe}},\ and\ \bibinfo {author} {\bibfnamefont {Y.}~\bibnamefont {Yanase}},\ }\bibfield  {title} {\bibinfo {title} {Theory of orbital magnetic quadrupole moment and magnetoelectric susceptibility},\ }\href {https://doi.org/10.1103/physrevb.98.020407} {\bibfield  {journal} {\bibinfo  {journal} {Phys. Rev. B}\ }\textbf {\bibinfo {volume} {98}},\ \bibinfo {pages} {020407} (\bibinfo {year} {2018})}\BibitemShut {NoStop}%
\end{thebibliography}
\end{document}